\documentclass[reprint,prc,twocolumn,amsmath,amssymb,aps,superscriptaddress,longbibliography,floatfix]{revtex4-2} %
\usepackage{graphicx}
\usepackage{tabularx}
\usepackage{float}
\usepackage{url}
\usepackage[switch]{lineno}
\usepackage[dvipsnames]{xcolor}
\usepackage{ragged2e}
\usepackage{enumitem}
\begin{document}
\preprint{APS/123-QED}

\title{First Results on Nucleon Resonance Electroexcitation Amplitudes from $ep\to e'\pi^+\pi^-p'$ Cross Sections at $W$ from 1.4-1.7 GeV and $Q^2$ from 2.0-5.0 GeV$^2$ }



\newcommand*{\JLAB}{Thomas Jefferson National Accelerator Facility, Newport News, Virginia 23606}
\newcommand*{\JLABindex}{43}
\affiliation{\JLAB}

\newcommand*{\SCAROLINA}{University of South Carolina, Columbia, South Carolina 29208}
\newcommand*{\SCAROLINAindex}{40}
\affiliation{\SCAROLINA}


\newcommand*{\ut}{Institute for Theoretical Physics, T\"ubingen University, D-72076 T\"ubingen, Germany}
\affiliation{\ut}

\newcommand*{\MSU}{Skobeltsyn Institute of Nuclear Physics and Physics Department, Lomonosov Moscow State University, 119234 Moscow, 
Russia}
\newcommand*{\MSUindex}{39}
\affiliation{\MSU}

\newcommand*{\UCONN}{University of Connecticut, Storrs, Connecticut 06269}
\newcommand*{\UCONNindex}{8}
\affiliation{\UCONN}

\author{V.I.~Mokeev}
\affiliation{\JLAB}
\author{P.~Achenbach}
\affiliation{\JLAB}
\author {V.D.~Burkert} 
\affiliation{\JLAB}
\author {D.S.~Carman} 
\affiliation{\JLAB}
\author {R.W.~Gothe} 
\affiliation{\SCAROLINA}
\author{A.N.~Hiller Blin}
\affiliation{\ut}
\author {E.L.~Isupov} 
\affiliation{\MSU}
\author {K.~Joo} 
\affiliation{\UCONN}
\author {K.~Neupane} 
\affiliation{\SCAROLINA}
\author {A.~Trivedi} 
\affiliation{\SCAROLINA}


\date{\today}

\begin{abstract}
The electroexcitation amplitudes or $\gamma_vpN^*$ electrocouplings of the $N(1440)1/2^+$, $N(1520)3/2^-$, and $\Delta(1600)3/2^+$
resonances were obtained for the first time from the $ep \to e'\pi^+\pi^-p'$ differential cross sections measured with the CLAS 
detector at Jefferson Lab within the range of invariant mass $W$ of the final state hadrons from 1.4--1.7~GeV for photon virtualities 
$Q^2$ from 2.0--5.0~GeV$^2$. A good description of the nine independent one-fold differential $\gamma_vp\to \pi^+\pi^-p'$ cross 
sections achieved within the data-driven JM meson-baryon reaction model in each bin of ($W$,$Q^2$) allows for separation of 
the resonant and non-resonant contributions. The electrocouplings were determined in independent fits of the $\pi^+\pi^-p$ cross 
sections within three overlapping $W$ intervals with a substantial contribution from each of the three resonances listed above. 
Consistent results on the electrocouplings extracted from the data in these $W$ intervals provide evidence for their reliable
extraction. These studies extend information on the electrocouplings of the $N(1440)1/2^+$ and $N(1520)3/2^-$ available from this 
channel over a broader range of $Q^2$. The electrocouplings of the $\Delta(1600)3/2^+$, which decays preferentially into $\pi\pi N$ 
final states, have been determined for the first time. Consistent results on the electrocouplings of the $N(1440)1/2^+$ and 
$N(1520)3/2^-$ from the $\pi N$ and $\pi^+\pi^-p$ channels allows for the determination of the uncertainties related to the reaction 
models employed in the data fits. The reliable extraction of the electrocouplings for these states is also supported by the
description of the $\pi^+\pi^-p$ differential cross sections with $Q^2$-independent masses and total/partial hadronic decay widths 
into the $\pi\Delta$ and $\rho p$ final states. Our results provide further evidence for the structure of these resonances in 
terms of an interplay between the inner core of three dressed quarks and an external meson-baryon cloud. 
\end{abstract}

\pacs{\bf{\textit{75.25.-j, 13.60.-r, 13.88.+e, 24.85.+p}} \vspace{-4pt}}

\keywords{Exclusive meson photo- and electroproduction, Charged double pion photo- and electroproduction, Structure of Nucleon resonances \vspace{-4pt}}

\maketitle


\section{Introduction}
\label{intro}

Studies of exclusive $\pi^+\pi^-p$ photo- and electroproduction off protons represent an effective tool for the exploration of the 
spectrum and structure of nucleon resonances~\cite{Mokeev:2022xfo,Carman:2020qmb, Brodsky:2020vco,Mokeev:2020hhu,Mokeev:2020vab}.
In this work we will use $N^*$ to represent both excited nucleon states $N$(mass)$J^P$, as well as Delta states $\Delta$(mass)$J^P$. 
The $\pi^0p$, $\pi^+n$, and $\pi^+\pi^-p$ electroproduction channels account for the largest part of the inclusive virtual 
photon-proton cross sections in the resonance excitation region \cite{Mokeev:2022xfo}. The data on $\pi N$ and $\pi^+\pi^-p$
electroproduction offer complementary information on the nucleon resonance electroexcitation amplitudes, the so-called
$\gamma_vpN^*$ electrocouplings, and their evolution with photon virtuality $Q^2$ ($Q^2 = -q_{\mu}^2$), where $q_{\mu}$ is the virtual
photon four-momentum. 

The low-lying $N^*$ states in the mass range below 1.6~GeV decay preferentially into $\pi N$, making single pion 
electroproduction data the driving source of information on the electrocouplings of these states~\cite{Aznauryan:2011qj,Park:2014yea,
Aznauryan:2009mx}. On the other hand, the branching fraction ($BF$) of these resonance decays into $\pi \pi N$ remains appreciable at 
the level of around 40\%, allowing for an independent determination of their electrocouplings from this channel
\cite{Mokeev:2012vsa,Mokeev:2015lda}. Consistent results on the electrocouplings from independent studies of $\pi N$ and 
$\pi^+\pi^-p$ demonstrate the capability of the $\pi^+\pi^-p$ reaction model to provide extraction of these 
quantities and to evaluate the systematic uncertainties associated with their determination~\cite{Aznauryan:2005tp,Mokeev:2015lda}.

Several $N^*$s in the mass range above 1.6~GeV decay preferentially into the $\pi \pi N$ final states with $BF$ around 70\%, making 
studies of $\pi^+\pi^-p$ electroproduction the major source of information on the electrocouplings of these states
\cite{Mokeev:2015lda}. At the same time, there are $N^*$s with masses above 1.6~GeV that decay mostly to the $\pi N$ 
final states~\cite{ParticleDataGroup:2022pth}. Therefore, studies of both $\pi N$ and $\pi^+\pi^-p$ electroproduction off protons are of
particular importance to get information on the $Q^2$-evolution of the electrocouplings for most prominent nucleon resonances. 

Coupled-channel approaches are making progress towards the extraction of the electrocouplings for low-lying $N^*$ states
from the combined analyses of meson photo-, electro-, and hadroproduction data. Recently, the $\pi N$ and $\eta p$ electroproduction 
multipoles, which are directly related to the $\gamma_vpN^*$ electrocouplings, were determined from CLAS data within a multi-channel
analysis~\cite{Mai:2021vsw,Mai:2021aui}. The first results on the electrocouplings of the $\Delta(1232)3/2^+$ and $N(1440)1/2^+$ at 
their pole positions in the complex energy plane have become available from the global multi-channel analysis developed by the
Argonne-Osaka Collaboration~\cite{Kamano:2018sfb}. The contributions from the $\pi\Delta$ and $\rho p$ channels deduced from the 
$\pi^+\pi^-p$ cross sections play an important role in the development of these approaches~\cite{Mokeev:2008iw,Mokeev:2015lda}.
Experiments of the 6-GeV era with the CLAS detector~\cite{Mecking:2003zu} in Hall~B at Jefferson Lab have provided the 
first and still only available information on the $Q^2$-evolution of the electrocouplings of most resonances in the mass range up to
1.8~GeV for $Q^2 < 5$~GeV$^2$~\cite{Mokeev:2022xfo,Carman:2020qmb}. 

Studies of $\pi^+\pi^-p$ photo- and electroproduction also provide a promising avenue in the search for the so-called ``missing"
resonances. Constituent quark models based on approximate symmetries of the strong interaction that are relevant for the strongly 
coupled regime, when the QCD running coupling $\alpha_s/\pi$ is comparable to unity, predict many more $N^*$s than have been seen 
in experiments with both electromagnetic and hadronic probes~\cite{Capstick:2000qj,Giannini:2015zia,Klempt:2009pi}. These expectations 
are supported by the results on the $N^*$ spectrum obtained starting from the QCD Lagrangian both within lattice and continuum QCD
approaches~\cite{Edwards:2011jj,Chen:2019fzn}.

The combined analysis of the CLAS $\pi^+\pi^-p$ photo- and electroproduction data~\cite{CLAS:2002xbv,CLAS:2018drk} carried out for 
$W$ from 1.6--1.8~GeV and $Q^2$ from 0--1.5~GeV$^2$ revealed the presence of a new $N'(1720)3/2^+$ baryon state~\cite{Mokeev:2020hhu}.
Only after implementation of this state with photo-/electrocouplings, mass, and decay widths fit to the CLAS data, was a successful 
description of the $\pi^+\pi^-p$ photo-/electroproduction data achieved with $Q^2$-independent masses and decay widths into the
$\pi\Delta$ and $\rho p$ final states. The contributions from the $N(1720)3/2^+$ and the new $N'(1720)3/2^+$ are well separated in 
the $\pi^+\pi^-p$ photo-/electroproduction data analyses despite their close masses and same spin-parity due to their different 
patterns for decay into intermediate $\pi \Delta$ and $\rho p$ states and the different $Q^2$-evolution of their electrocouplings. 
These differences can only be seen in the combined studies of $\pi^+\pi^-p$ photo- and electroproduction, but they were elusive 
in previous studies of the two-body meson-baryon channels. Therefore, such combined studies provide a promising avenue in the quest 
to discover additional resonances.   

The CLAS results on the electrocouplings make it possible for the first time to determine the resonant contributions to the inclusive
electron scattering structure functions in the resonance region~\cite{HillerBlin:2022ltm,Blin:2021twt,HillerBlin:2019jgp}. They
also provide a new opportunity to better understand the ground state nucleon parton distribution functions (PDFs) at large values of 
the fractional parton momentum $x$ within the resonance region.

Analyses of the CLAS results on the $Q^2$-evolution of the electrocouplings within coupled-channel approaches
\cite{Kamano:2018sfb,Suzuki:2010yn} and continuum Schwinger methods (CSMs)~\cite{Burkert:2017djo}, supported by the results from 
different quark models~\cite{Aznauryan:2014xea,Aznauryan:2018okk,Obukhovsky:2019xrs,Lyubovitskij:2020gjz}, have revealed $N^*$ 
structure as an interplay between the inner core of three dressed quarks and an external meson-baryon cloud
\cite{Mokeev:2022xfo,Brodsky:2020vco,Barabanov:2020jvn,Kamano:2018sfb}. Studies of the electrocouplings for $Q^2 \lesssim 2$~GeV$^2$ 
provide important information on the transition from confined dressed quarks to deconfined mesons and baryons that give 
rise to the meson-baryon cloud.

Analyses of these electrocouplings suggest a gradual transition from the convolution between the meson-baryon cloud and quark core in 
$N^*$ structure towards quark core dominance with increasing $Q^2$~\cite{Burkert:2017djo,Aznauryan:2018okk,Obukhovsky:2019xrs,
Lyubovitskij:2020gjz}. Virtual photons with $Q^2 \gtrsim 1-2$~GeV$^2$ penetrate the meson-baryon cloud and interact mostly with the 
quark core. Consequently, studies of the electrocouplings in this higher $Q^2$ regime provide a unique way to explore the structure 
of dressed quarks and the evolution of their interactions at distance scales from the strongly coupled to the perturbative (pQCD)
regimes~\cite{Carman:2023zke,Roberts:2021xnz,Rodriguez-Quintero:2019yec,Roberts:2020hiw}. Therefore, the region of high $Q^2$ looks
promising to explore many facets of the strong interaction dynamics between three dressed quarks apparent in the generation of 
various $N^*$s with different quantum numbers and structure.

The description of the $\Delta(1232)3/2^+$ and $N(1440)1/2^+$ electrocouplings achieved within CSMs~\cite{Segovia:2014aza,
Segovia:2015hra} by employing the same momentum dependence of the dressed quark mass deduced from the QCD Lagrangian
\cite{Roberts:2020hiw,Roberts:2021xnz} and used in the description of the pion and nucleon elastic electromagnetic form factors
\cite{Roberts:2021nhw, Horn:2016rip,Segovia:2014aza} demonstrated the promising opportunity for gaining insight into the emergence of 
more than 98\% of the hadron mass from data on the $Q^2$-evolution of the electrocouplings.

As of now, the electrocouplings are available from $\pi^+\pi^-p$ electroproduction cross sections within the range 
of $Q^2 < 1.5$~GeV$^2$. In this paper, we present an extension of the results on the electrocouplings of the $N(1440)1/2^+$ and
$N(1520)3/2^-$ determined from the $\pi^+\pi^-p$ cross sections at $W$ from 1.4--1.7~GeV for $Q^2$ from 2.0--5.0~GeV$^2$ and compare 
them with the available results from the studies of $\pi N$ electroproduction within the same kinematic domain. 
The $\Delta(1600)3/2^+$ recently was elevated to a four-star PDG status~\cite{ParticleDataGroup:2022pth}. This state decays
preferentially into $\pi \pi N$. The electrocouplings of the $\Delta(1600)3/2^+$ have become available for the first time from the 
analysis of the $\pi^+\pi^-p$ electroproduction data presented here. 

This paper is organized as follows. In Section~\ref{data_model} we present the kinematic variables for the description of $\pi^+\pi^-p$
electroproduction and the one-fold differential cross sections measured with CLAS used for the extraction of the electrocouplings in the
mass range below 1.7~GeV. Here, we also discuss the updates to the Jefferson Lab-Moscow State University (JM) reaction model relevant 
to the extraction of the electrocouplings for $Q^2$ from 2.0--5.0~GeV$^2$. The procedures developed for the evaluation of the
electrocouplings from the cross section fits are presented in Section~\ref{fit_strategy}. The results on the electrocouplings and 
partial decay widths to $\pi\Delta$ and $\rho p$ for the $N(1440)1/2^+$, $N(1520)3/2^-$, and $\Delta(1600)3/2^+$ are presented in
Section~\ref{elcoupl_hadrdec}, along with comparisons of the electrocouplings from the $\pi^+\pi^-p$ data with those previously 
available from the analyses of $\pi N$ data. The impact of these results on the understanding of $N^*$ structure is presented in 
Section~\ref{impact_theor}. We conclude and highlight the future prospects for resonance electrocoupling studies from exclusive meson
electroproduction data in Section~\ref{concl_outlook}.

\section{Cross Sections and Reaction Model for Electrocouplings}
\label{data_model}

In this section we describe the $\pi^+\pi^-p$ differential cross sections measured with CLAS for $W$ from 1.4--1.7~GeV and $Q^2$ from 
2.0--5.0~GeV$^2$~\cite{CLAS:2017fja,Trivedi:2018rgo} that were used for the extraction of the $N(1440)1/2^+$, 
$N(1520)3/2^-$, and $\Delta(1600)3/2^+$ parameters. We also present the basic features of the JM model relevant for the extraction of 
the electrocouplings from the $\pi^+\pi^-p$ data and the most recent JM model updates.

\begin{figure}[htbp]
\begin{center}
\includegraphics[width=8cm]{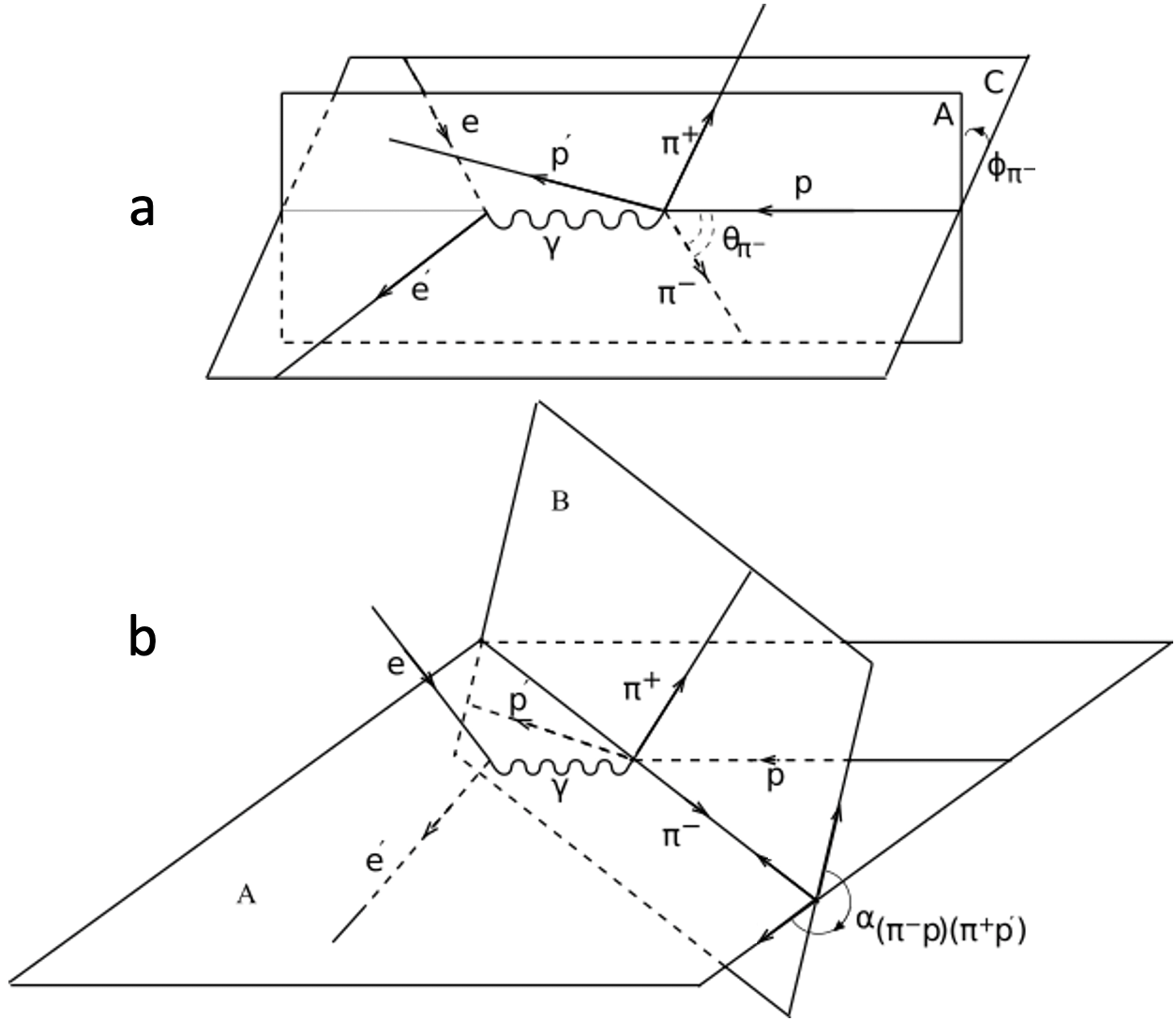}
\caption{Kinematic variables for the description of the reaction $\gamma_v p \to  \pi^+ \pi^- p'$ in the CM frame of the final state
hadrons corresponding to the $d^5\tau$ differential assignment given in Section~\ref{kinematics}. Panel (a) shows the $\pi^-$ polar 
and azimuthal angles $\theta_{\pi^-}$ and $\phi_{\pi^-}$. Plane C represents the electron scattering plane. The $z$-axis is 
directed along the $\gamma_v$ three-momentum, while the $x$-axis is located in the electron scattering plane C and the $y$-axis 
forms a right-handed coordinate system. Plane A is defined by the three-momenta of the initial state proton and the final state 
$\pi^-$. Panel (b) shows the angle $\alpha_{[\pi^-p][\pi^+p']}$ between the two hadronic planes A and B or the plane B rotation angle 
around the axis aligned along the three-momentum of the final state $\pi^-$. Plane B is defined by the three-momenta of the final state 
$\pi^+$ and $p'$.} 
\label{fig_kinematic}
\end{center}
\end{figure}

\begin{table}
\begin{center}
\begin{tabular}{|c|c|} \hline
                         & 2.0-2.4    \\
                         & 2.2 for computed cross sections \\ \cline{2-2}
                         & 2.4-3.0   \\
                         & 2.6 for computed cross sections \\ \cline{2-2}
                         & 3.0-3.5   \\
$Q^2$ Interval, GeV$^2$ & 3.2 for computed cross sections \\ \cline{2-2}
                         & 3.5-4.2    \\
                         & 3.6 for computed cross sections \\  \cline{2-2}
                         & 4.2-5.0   \\
                         & 4.4 for computed cross sections \\  \hline
$W$ Interval, GeV       & 1.41-1.66 \\
covered in each $Q^2$ bin & 11 bins   \\ \hline
\end{tabular}
\caption{Kinematic area covered in the fit of the CLAS $\pi^+\pi^-p$ electroproduction cross sections for the extraction of the 
resonance parameters~\cite{CLAS:2017fja,Trivedi:2018rgo}.}
\label{wq2bins} 
\end{center}
\end{table}

\begin{table}
\begin{center}
\begin{tabular}{|c|c|c|}
\hline
One-Fold Differential & Interval  & Number of   \\
Cross Section         & Covered   & Bins        \\ \hline
$\frac{d\sigma}{dM_{\pi^+p}}$ ($\mu$b/GeV)         & $M_{\pi^+p}^{min}$-$M_{\pi^+p}^{max}$ & 14  \\
$\frac{d\sigma}{dM_{\pi^+\pi^-}}$ ($\mu$b/GeV)     & $M_{\pi^+\pi^-}^{min}$-$M_{\pi^+\pi^-}^{max}$ & 14  \\
$\frac{d\sigma}{dM_{\pi^-p}}$ ($\mu$b/GeV)         & $M_{\pi^-p}^{min}$-$M_{\pi^-p}^{max}$ & 14  \\
$\frac{d\sigma}{\sin \theta_{\pi^-} d\theta_{\pi^-}}$ ($\mu$b/rad)      & 0-180$^\circ$ & 10  \\
$\frac{d\sigma}{\sin \theta_{\pi^+} d\theta_{\pi^+}}$ ($\mu$b/rad)      & 0-180$^\circ$ & 10  \\
$\frac{d\sigma}{\sin \theta_{p'}d\theta_{p'}}$ ($\mu$b/rad)        & 0-180$^\circ$ & 10  \\
$d\sigma/d\alpha_{[\pi^-p][\pi^+p']}$ ($\mu$b/rad) & 0-360$^\circ$ & 10\\
$d\sigma/d\alpha_{[\pi^+p][\pi^-p']}$ ($\mu$b/rad) & 0-360$^\circ$ & 10\\
$d\sigma/d\alpha_{[\pi^+\pi^-][p p']}$ ($\mu$b/rad)& 0-360$^\circ$ & 10\\ \hline
\end{tabular}
\caption{List of the one-fold differential cross sections measured with CLAS~\cite{CLAS:2017fja,Trivedi:2018rgo} 
and the binning over the kinematic variables. $M_{i,j}^{min}=M_i+M_j$ and $M_{i,j}^{max} = W-M_k$, where $M_{i,j}$ and $M_k$ are the 
invariant masses of the final state hadron pair $(i,j)$, and the mass of the third final state hadron $k$, respectively.}
\label{1diffbins}
\end{center}
\end{table}

\subsection{Kinematic Variables and $\pi^+\pi^-p$ Electroproduction Cross Sections}
\label{kinematics}

For the $\gamma_v p \to \pi^+\pi^-p'$ reaction, the invariant mass of the final state hadrons $W$ and photon virtuality $Q^2$
unambiguously determine the initial state virtual photon and proton four-momenta in their center-of-mass (CM) frame with the 
$z$-axis directed along the $\gamma_v$ three-momentum as shown in Fig.~\ref{fig_kinematic}. The final $\pi^+\pi^-p$ state is 
described by the four-momenta of the three final state hadrons by twelve variables. Energy-momentum conservation reduces the number 
of variables down to eight. Since the three final state hadrons are on-shell, three additional relations between the final state 
hadron energies and absolute momentum values reduce the number of independent variables down to five. Hence, at a given $W$ and 
$Q^2$, the reaction can be fully described by the five-fold differential cross section $d^5\sigma/d^5\tau$, where $d^5\tau$ is
differential in the five independent variables that determine the final state hadron four-momenta. There are many choices for 
these five variables~\cite{Byckling:1971vca}. After defining $M_{\pi^+p}$, $M_{\pi^-p}$, and $M_{\pi^+\pi^-}$ as the invariant 
masses of the three possible two-hadron pairs in the final state, we adopt here the following assignment for the computation of 
the five-fold differential cross section: $d^5\tau = dM_{\pi^+p} dM_{\pi^+\pi^-} d\Omega_{\pi^-} d\alpha_{[\pi^-p][\pi^+p']}$, 
where $\Omega_{\pi^-}$ is the final state $\pi^-$ solid angle defined by the polar ($\theta_{\pi^-}$) and azimuthal ($\phi_{\pi^-}$) 
angles shown in Fig.~\ref{fig_kinematic}(a), and $\alpha_{[\pi^-p][\pi^+p']}$ is the rotation angle of plane B defined by the momenta 
of the final state $\pi^+$ and $p'$ around the axis defined by the final state $\pi^-$ momentum, see Fig.~\ref{fig_kinematic}(b). This
$d^5\tau$ differential is used in the computation of the $\pi^+\pi^-p$ cross sections within the JM model for comparison with the 
experimental data~\cite{CLAS:2018fon,CLAS:2008ihz}. All frame-dependent variables are defined in the final state hadron CM frame.

The $\pi^+\pi^-p$ electroproduction data have been collected in the bins of a seven-dimensional space, since for the
description of the initial state kinematics, $W$ and $Q^2$ are required. The number of bins in the seven-dimensional reaction phase 
space and the kinematic area covered by the data for extraction of the differential cross sections are detailed in Tables~\ref{wq2bins}
and \ref{1diffbins}. The huge number of seven-dimensional bins over the reaction phase space ($\approx 1 \times 10^7$ bins) does not 
allow us to use the correlated multi-fold differential cross sections in the analysis of the data. More than half of the 
five-dimensional phase-space bins of the final state hadrons at any given $W$ and $Q^2$ remain unpopulated due to statistical
limitations. Therefore, we use the following one-fold differential cross sections in each bin of $W$ and $Q^2$ covered by the data:

\begin{itemize}
\item invariant mass distributions for the three pairs of the final state particles $d\sigma/dM_{\pi^+\pi^-} $, $d\sigma/dM_{\pi^+ p}$, 
and $d\sigma/dM_{\pi^- p}$;
\item distributions for the CM polar angles of the three final state particles $d\sigma/(\sin \theta_{\pi^-} d\theta_{\pi^-})$, 
$d\sigma/(\sin \theta_{\pi^+} d\theta_{\pi^+})$, and $d\sigma/(\sin \theta_{p'} d\theta_{p'})$;
\item distributions for the three $\alpha$-angles determined in the CM frame: $d\sigma/d\alpha_{[\pi^-p][\pi^+p']}$,
$d\sigma/d\alpha_{[\pi^+p][\pi^-p']}$, and $d\sigma/d\alpha_{[\pi^+\pi^-][p p']}$, where $d\sigma/d\alpha_{[\pi^+p][\pi^-p']}$ and
$d\sigma/d\alpha_{[\pi^+\pi^-][p p']}$ are defined analogously to $d\sigma/d\alpha_{[\pi^-p][\pi^+p']}$
described above. 
\end{itemize} 

The one-fold differential cross sections were obtained by integrating the five-fold differential cross sections over the other four
kinematic variables of $d^5\tau$. However, the angular distributions for the polar angles of the final state $\pi^+$ and $p$, as well 
as for the rotation angles around the axes along the momenta of these final state hadrons, cannot be obtained from $d^5\tau$
described above, since this differential does not depend on these variables. Two other sets of differentials $d^5\tau'$ and $d^5\tau''$ 
are required, which contain $d\Omega_{\pi^+}d\alpha_{[\pi^+p][\pi^-p']}$ and $d\Omega_{p'}d\alpha_{[pp'][\pi^+\pi^-]}$, respectively, as 
described in Refs.~\cite{CLAS:2008ihz,Mokeev:2012vsa}. The five-fold differential cross sections evaluated over the other two 
differentials were computed from the five-fold differential cross section over the $d^5\tau$ differential by means of cross section 
interpolation. For each kinematic point in the five-dimensional phase space determined by the variables of the $d^5\tau'$ and 
$d^5\tau''$ differentials, the four-momenta of the three final state hadrons were computed, and from these values, the five variables 
of the $d^5\tau$ differential were determined. The $d^5\sigma/d^5\tau$ cross sections were interpolated into this five-dimensional 
kinematic point.

\subsection{Reaction Model for Extraction of Electrocouplings}
\label{react_model}

The $N(1440)1/2^+$, $N(1520)3/2^-$, and $\Delta(1600)3/2^+$ electrocouplings have been extracted for $Q^2$ from
2.0--5.0~GeV$^2$ from the data on the $\pi^+\pi^-p$ differential cross sections by fitting them within the framework of the 
data-driven JM reaction model detailed in Refs.~\cite{Mokeev:2008iw,Mokeev:2012vsa,Mokeev:2015lda}, referred to as JM17, which was 
used for the extraction of the electrocouplings for $Q^2 < 1.5$~GeV$^2$ and $W < 1.8$~ GeV. Within this
approach, the $\pi^+\pi^-p$ electroproduction mechanisms seen through their manifestation in the observables as peaks in
the invariant mass distributions for the final state hadrons and with pronounced dependencies in the CM angular distributions for the 
final state hadrons were incorporated. The remaining mechanisms without pronounced kinematic dependencies were accounted for by 
exploring the correlations between the shapes of their contributions into the nine independent one-fold differential cross sections.

\begin{figure*}[htp]
\begin{center}
\includegraphics[width=13.8cm]{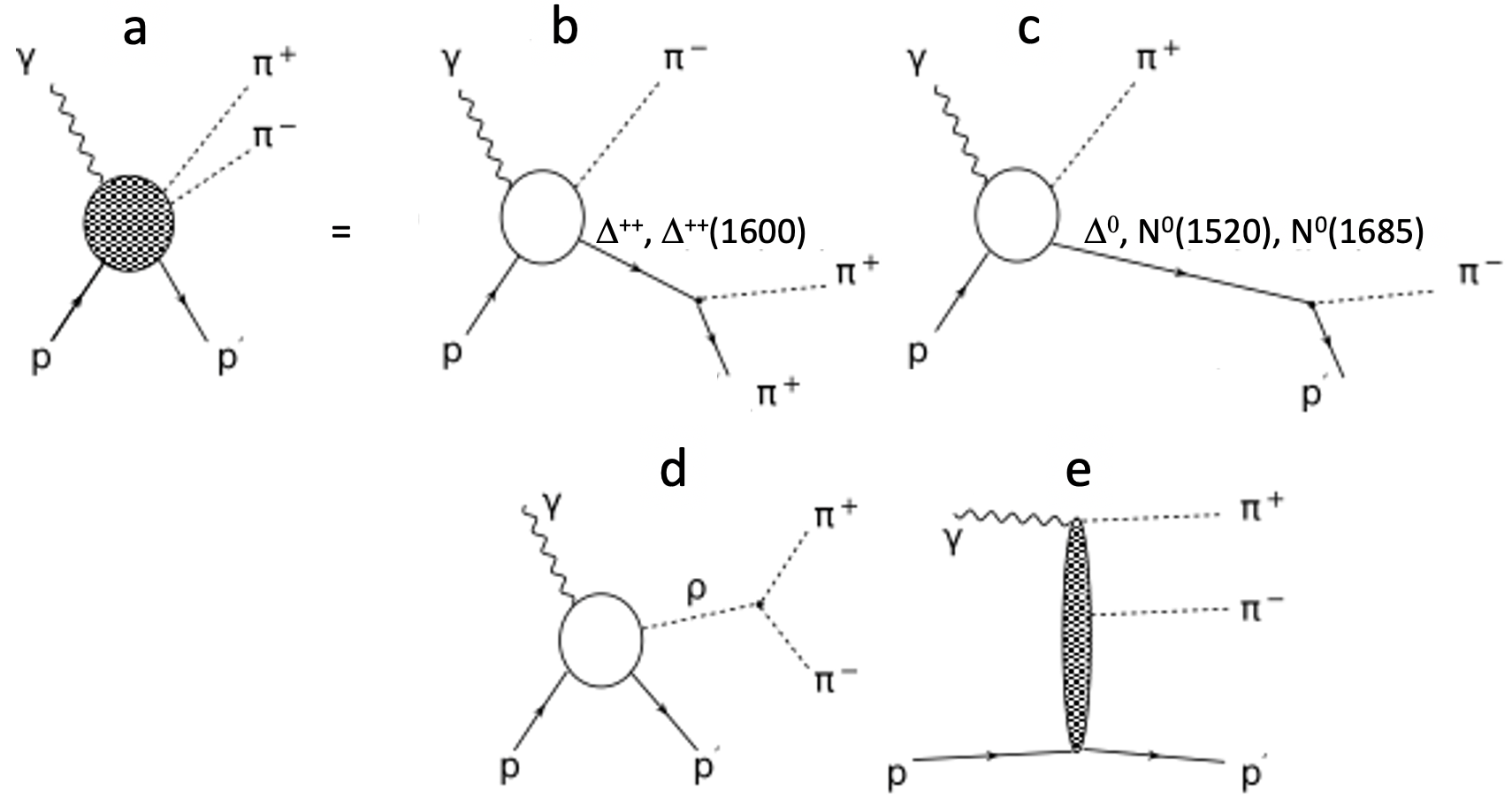}
\caption{The $\gamma_v p\to  \pi^+\pi^- p'$ electroproduction mechanisms incorporated at the amplitude level into the JM17 model
\cite{Mokeev:2008iw,Mokeev:2012vsa,Mokeev:2015lda}: a) full amplitude; b) $\pi^- \Delta^{++}$ and $\pi^-\Delta^{++}(1600)3/2^+$ 
sub-channels; c) $\pi^+ \Delta^0$, $\pi^+ N^0(1520)3/2^-$, and $\pi^+N^0(1680)5/2^+$ sub-channels; d) $\rho p$ sub-channel; e)
direct $2 \pi$ mechanisms.}
\label{jmmech}
\end{center}
\end{figure*}

The mechanisms incorporated into the JM model are shown in Fig.~\ref{jmmech}. The amplitudes of  the $\gamma_v p \to \pi^+\pi^- p'$ 
reaction are described as a superposition of the $\pi^-\Delta^{++}$, $\pi^+\Delta^0$, $\rho p$, $\pi^+ N^0(1520)$, and 
$\pi^+ N^0(1680)$ sub-channels with subsequent decays of the unstable hadrons to the final state $\pi^+\pi^-p$ state as detailed in 
Appendix III of Ref.~\cite{Mokeev:2008iw}. In addition, direct $2 \pi$ production mechanisms, where the final $\pi^+\pi^- p$  
comes about without going through the intermediate process of forming unstable hadron states are included. Evidence for these
contributions is seen in analyses of the final state hadron angular distributions with the phenomenological amplitudes described in
Ref.~\cite{Mokeev:2008iw}. 

\begin{table}
\begin{center}
\begin{tabular}{|c|c|c|c|} \hline
$N^*$ States & Mass,\     & Total  Decay        & Refs.  \\
Incorporated &  GeV       & Width               &     \\
in Data Fit  &            & $\Gamma_{tot}$, GeV &     \\ \hline
$N(1440)1/2^+$ & 1.43-1.48 & 0.25-0.40 & \cite{Aznauryan:2009mx}  \\
$N(1520)3/2^-$ & 1.51-1.53 & 0.12-0.13 & \cite{Aznauryan:2009mx}  \\
$N(1535)1/2^-$ & 1.51-1.55 & 0.12-0.18 & \cite{Aznauryan:2009mx} \\
$N(1650)1/2^-$ & 1.64-1.67 & 0.15-0.16 & \cite{HillerBlin:2019jgp}  \\
$N(1680)5/2^+$ & 1.68-1.69 & 0.11-0.13 & \cite{Park:2014yea}  \\
$N(1700)3/2^-$ & 1.65-1.75 & 0.16-0.18 & \cite{Park:2014yea} \\
$N'(1720)3/2^+$ & 1.71-1.74 & 0.11-0.13 &\cite{Mokeev:2020hhu}  \\
$N(1720)3/2^+$ & 1.73-1.76 & 0.11-0.13 & \cite{HillerBlin:2019jgp}  \\ 
$\Delta(1600)3/2^+$ & 1.50-1.64 & 0.20-0.30 &\cite{Lu:2019bjs}   \\
$\Delta(1620)1/2^-$ & 1.60-1.66 & 0.11-0.15 &\cite{HillerBlin:2019jgp}   \\
$\Delta(1700)3/2^-$ & 1.67-1.73 & 0.23-0.32 &\cite{HillerBlin:2019jgp}   \\ \hline
\end{tabular}
\caption{List of resonances included in the fit of the $\pi^+\pi^-p$ differential cross sections within the JM23 model and their 
parameters: masses, total decay widths $\Gamma_{tot}$, and ranges of their variation. The JM17 model contains all listed resonances,
except for the $\Delta(1600)3/2^+$. The starting values for the resonance electrocouplings were taken from the references given in 
the last column. The electrocouplings of the $N(1650)1/2^-$, $N(1720)3/2^+$, $N'(1720)3/2^+$, and $\Delta(1620)1/2^-$ were obtained 
for $Q^2 < 1.5$~GeV$^2$ and extrapolated to the $Q^2$ area covered by the CLAS data~\cite{CLAS:2017fja,Trivedi:2018rgo} as described 
in Ref.~\cite{HillerBlin:2019jgp}. The predictions of Ref.~\cite{Lu:2019bjs} were used as the starting values for the 
$\Delta(1600)3/2^+$ electrocouplings.}
\label{nstlist} 
\end{center}
\end{table}

Within the JM17 model, only the $\pi^-\Delta^{++}$, $\pi^+\Delta^0$, and $\rho p$ channels contain contributions from $N^*$s
excited in the $s$-channel for the $\gamma_v p$ interaction. The JM17 model incorporates contributions from all well-established 
$N^*$ states listed in Table~\ref{nstlist}, except for the $\Delta(1600)3/2^+$ that was not included. Note that the four-star 
$N(1675)5/2^-$ and $N(1710)1/2^+$ states were not included. The amplitudes for electroexcitation of the $N(1675)5/2^-$ 
off protons are suppressed in comparison with the electrocouplings of other resonances in the third resonance region. Furthermore, 
in this work we have only determined the $\gamma_vpN^*$ electrocouplings for resonances in the mass range below 1.6~GeV. In this 
case, only the tail from the weakly excited $N(1675)5/2^-$ can contribute. As well, studies of $\pi^+\pi^-p$ electroproduction in 
the third resonance region have revealed no evidence for contributions from the $N(1710)1/2^+$\cite{Mokeev:2020hhu,CLAS:2002xbv}. 

The resonant amplitudes are described by a unitarized Breit-Wigner ansatz~\cite{Mokeev:2012vsa}, which accounts for the transition
between the same and different resonances in the dressed resonance propagator, which makes the resonant amplitudes consistent with 
restrictions imposed by a general unitarity condition~\cite{Aitchison:1972ay,Kamano:2008gr}. Quantum number conservation in the 
strong interaction allows for transitions between the pairs of $N^*$ states, $N(1520)3/2^- \leftrightarrow N(1700)3/2^-$, 
$N(1535)1/2^- \leftrightarrow N(1650)1/2^-$, and $N(1720)3/2^+ \leftrightarrow N'(1720)3/2^+$, which are incorporated into the JM17 
model.

The non-resonant amplitudes in the $\pi\Delta$ sub-channels are described by the minimal set of current conserving 
Reggeized Born terms detailed in Ref.~\cite{Mokeev:2008iw}. They include the contact, Reggeized $t$-channel $\pi$-in-flight, 
$s$-channel nucleon, and $u$-channel $\Delta$-in-flight terms. Note, as $W$ is going to threshold, the Reggeized $t$-channel
term gradually transforms into the $\pi$-pole term, allowing use of the Reggeized Born terms at low $W$.

\begin{figure*}[htbp]
\begin{center}
\includegraphics[width=7.1cm]{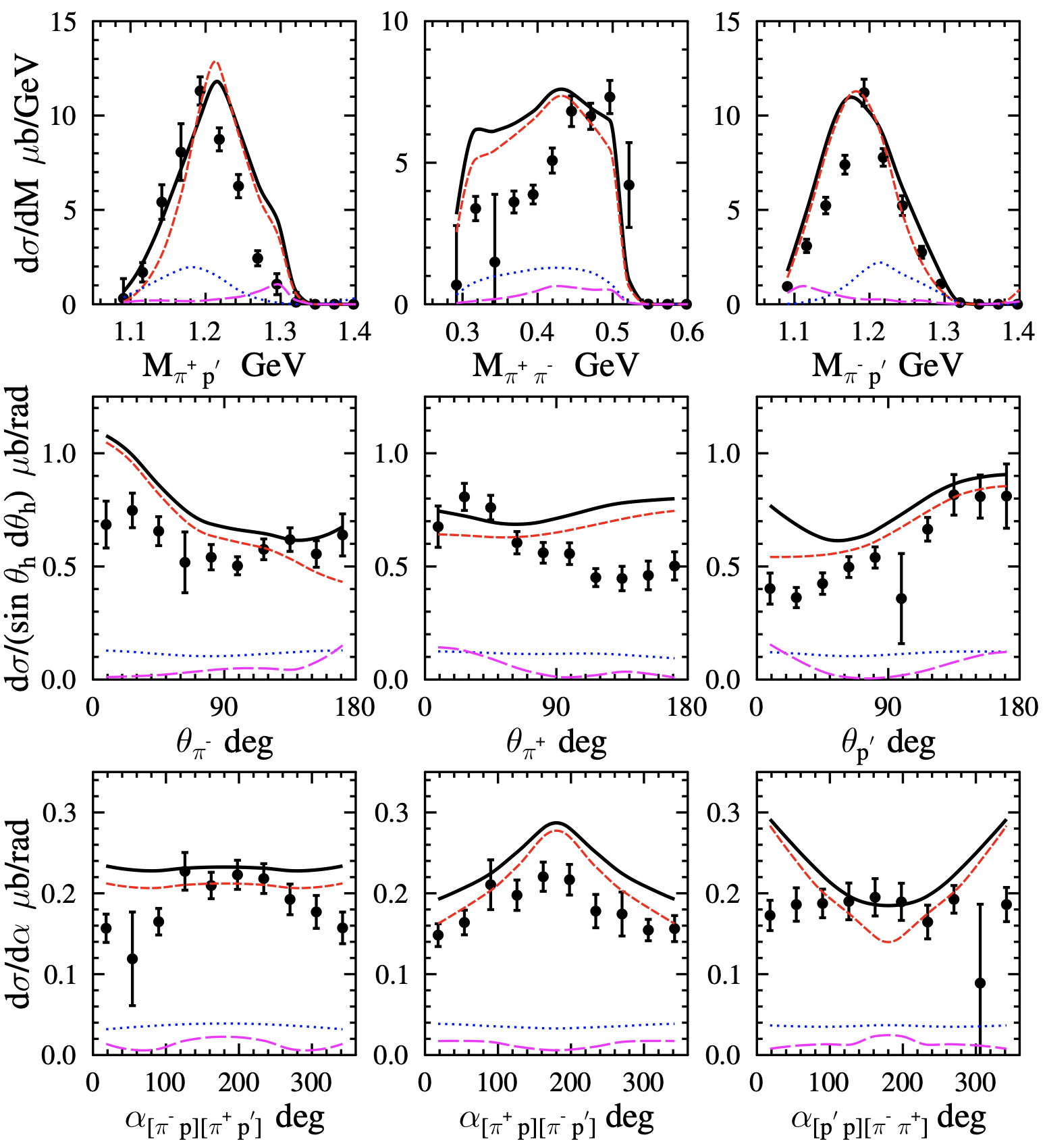}
\hspace{5mm}
\raisebox{0.8mm}{\includegraphics[width=7.1cm]{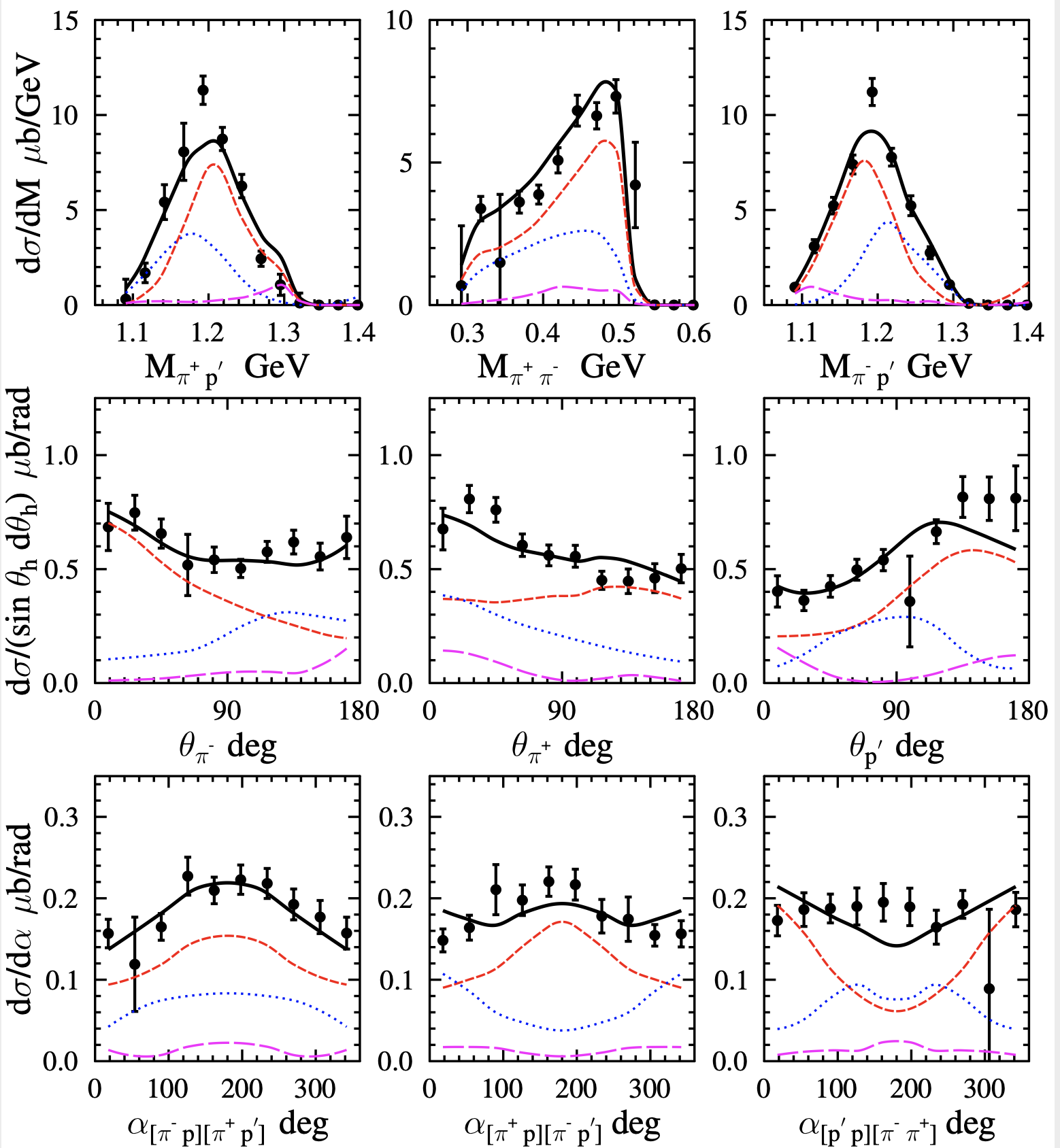}}
\caption{Description of the nine one-fold differential $\pi^+\pi^-p$ electroproduction cross sections measured with CLAS
\cite{CLAS:2017fja,Trivedi:2018rgo} (black points) achieved within the JM17 reaction model (left) and after improvements within the
updated version JM23 (right) at $W$ from 1.45--1.48~GeV and $Q^2$ from 3.50--4.20 GeV$^2$. $\theta_h$ represents the CM emission 
angles of the final state hadrons ($h = \pi^-, \pi^+, p$). The computed differential cross sections are shown by the black solid 
lines, while the contributions from the $\pi^-\Delta^{++}$ and $\pi^+\Delta^0$ channels are shown by the red dashed and blue dotted
lines, respectively. The contributions from direct $2 \pi$ production are shown by the magenta long-dashed lines.} 
\label{fig_jm17jm23_370146}
\end{center}
\end{figure*}

\begin{figure*}[htbp]
\begin{center}
\includegraphics[width=7.1cm]{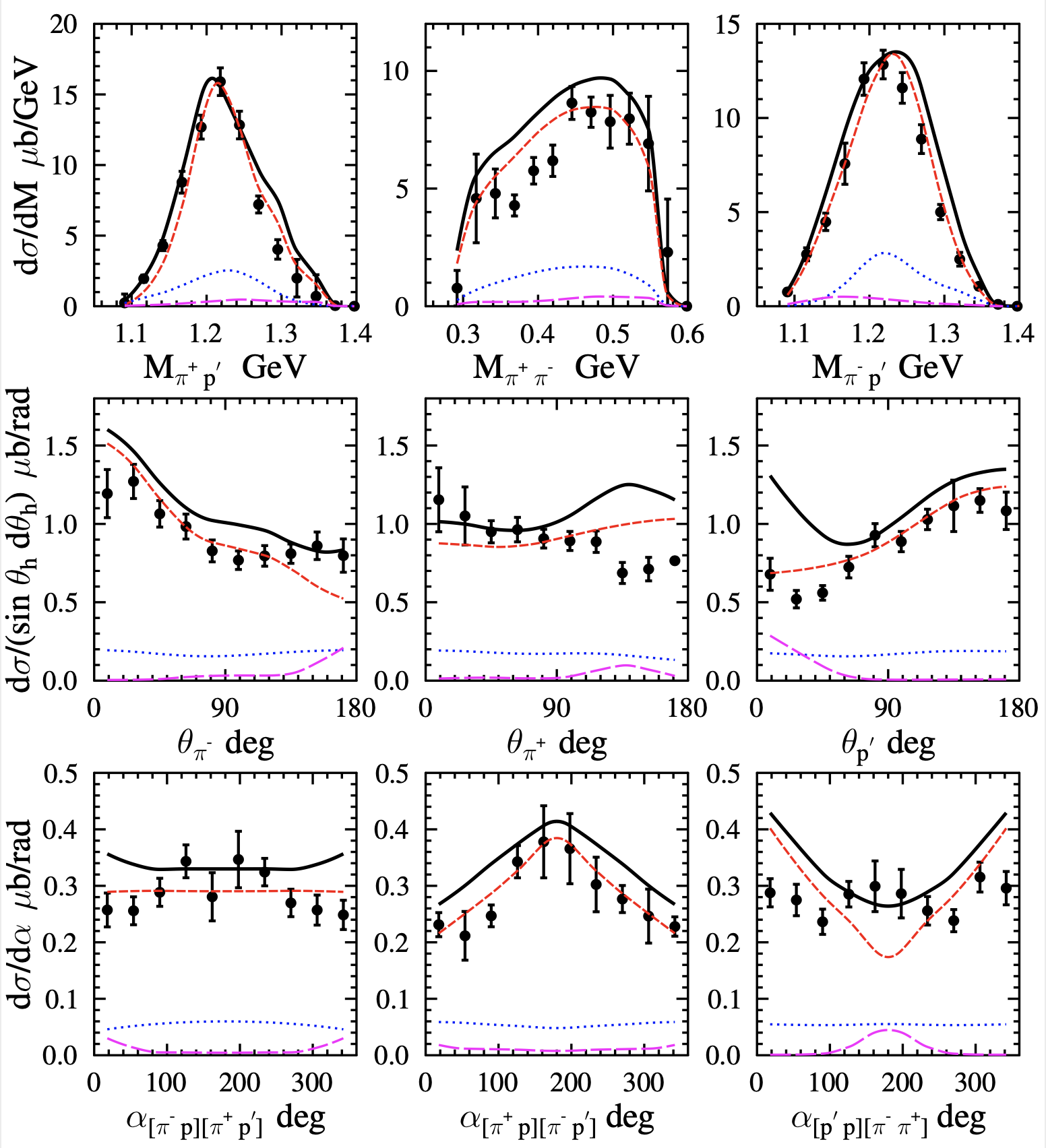}
\hspace{5mm}
\raisebox{0.2mm}{\includegraphics[width=7.1cm]{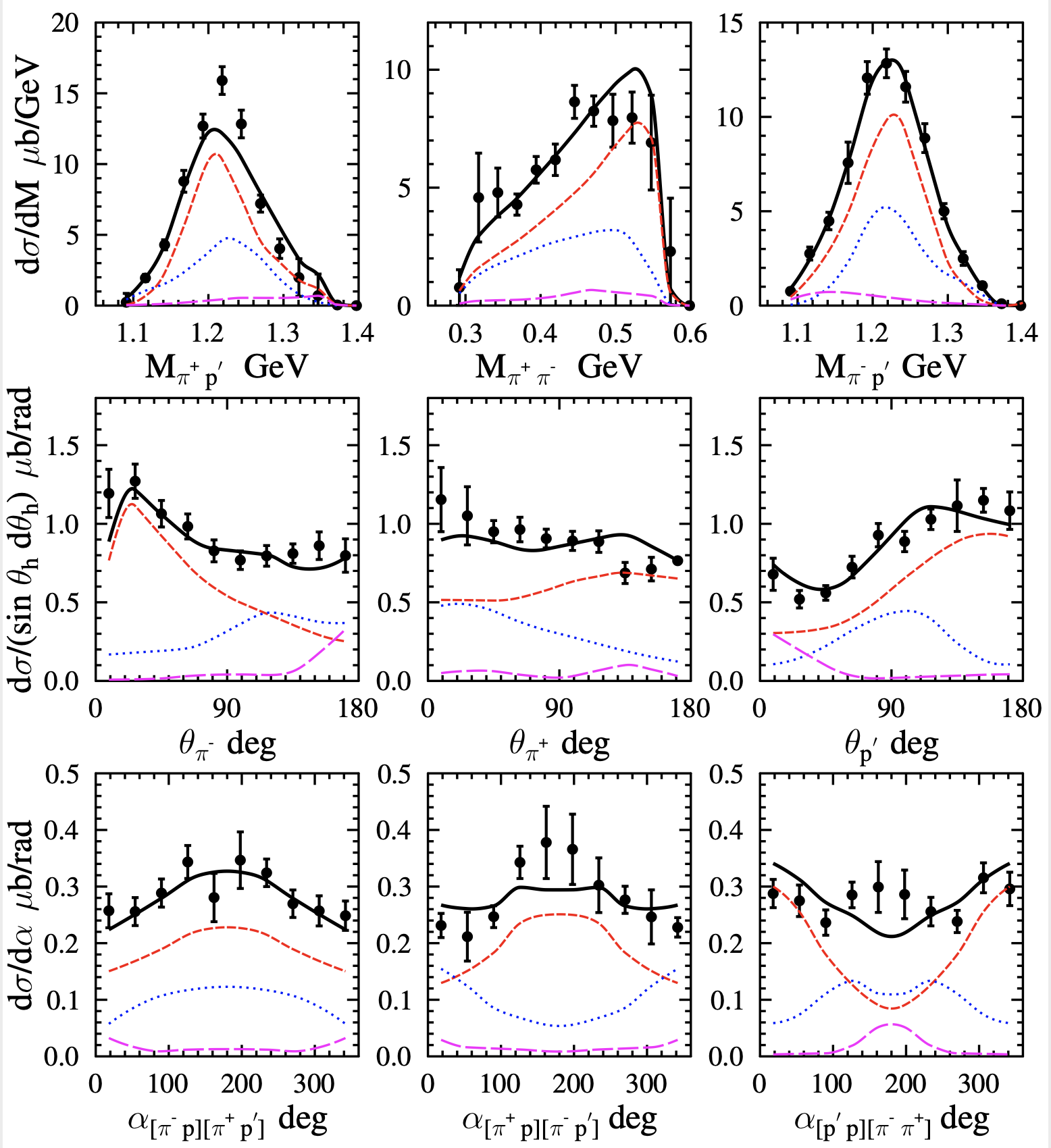}}
\caption{Description of the nine one-fold differential $\pi^+\pi^-p$ electroproduction cross sections measured with CLAS 
\cite{CLAS:2017fja,Trivedi:2018rgo} (in red) achieved within the JM17 reaction model (left) and after improvements 
within the updated version JM23 (right) at $W$ from 1.50--1.53~GeV and $Q^2$ from 3.50--4.20~GeV$^2$. The legend for the curves is 
the same as in Fig.~\ref{fig_jm17jm23_370146}.} 
\label{fig_jm17jm23_3701451}
\end{center}
\end{figure*}

\begin{figure*}[htbp]
\begin{center}
\includegraphics[width=7.1cm]{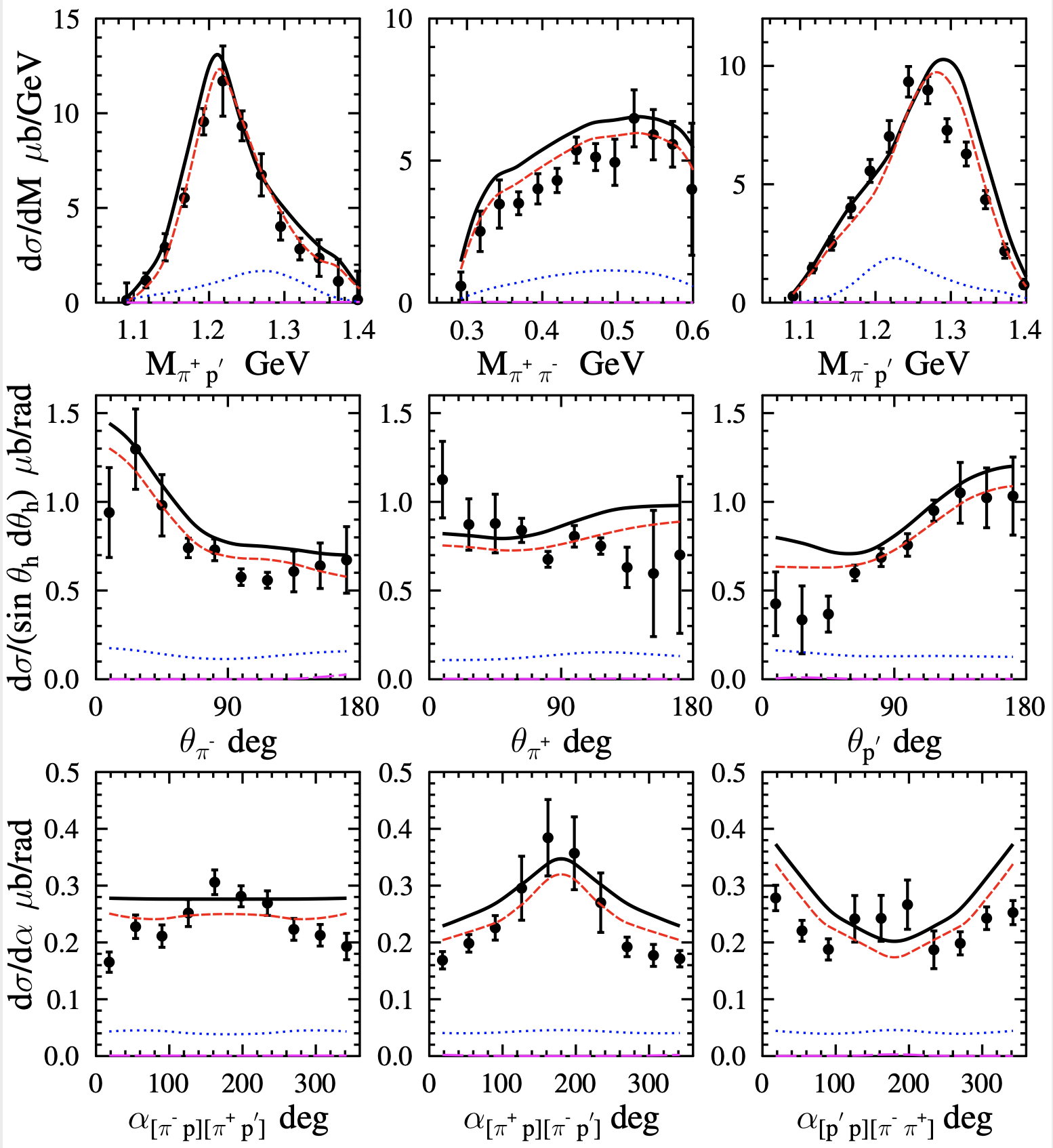}
\hspace{5mm}
\includegraphics[width=7.1cm]{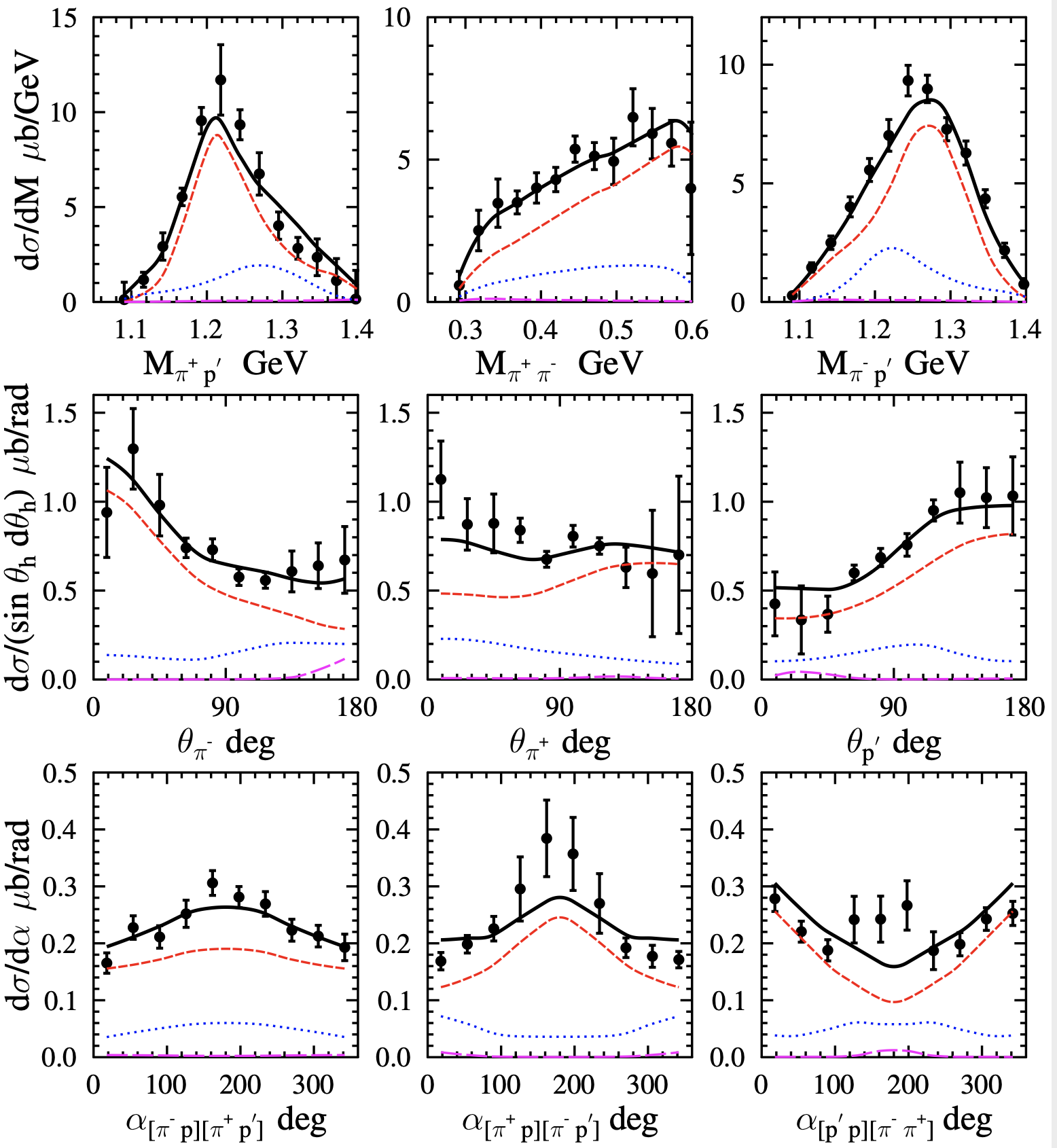}
\caption{Description of the nine one-fold differential $\pi^+\pi^-p$ electroproduction cross sections measured with CLAS 
\cite{CLAS:2017fja,Trivedi:2018rgo} (in red) achieved within the JM17 reaction model (left) and after improvements 
within the updated version JM23 (right) at $W$ from 1.55--1.58~GeV and $Q^2$ from 3.50--4.20~GeV$^2$. The legend for the curves is 
the same as in Fig.~\ref{fig_jm17jm23_370146}.} 
\label{fig_jm17jm23_3701456}
\end{center}
\end{figure*}

The $\pi^+ N^0(1520)3/2^-$ and $\pi^+ N^0(1680)5/2^+$ channels are described in the JM17 model by non-resonant contributions only. 
The amplitudes of the $\pi^+ N^0(1520)3/2^-$ sub-channel were derived from the non-resonant Born terms in the $\pi \Delta$ sub-channels 
by implementing an additional $\gamma_5$-matrix  that accounts for the opposite parities of the $\Delta(1232)3/2^+$ and $N(1520)3/2^-$
\cite{Mokeev:2008iw}. The magnitudes of the $\pi^+ N^0(1520)3/2^-$ production amplitudes were independently fit to the data for each 
bin in $W$ and $Q^2$. The contributions from the $\pi^+ N^0(1520)3/2^-$ sub-channel should be taken into account for $W > 1.5$~GeV.
The $\pi^+ N^0(1680)5/2^+$  contributions are seen in the data at $W > 1.6$~GeV. These contributions are almost negligible at smaller 
$W$. Effective $t$-channel exchange terms were employed in the JM17 model for parameterization of the amplitudes of this sub-channel
\cite{Mokeev:2008iw}. The magnitudes of the $\pi^+ N^0(1680)5/2^+$  amplitudes were fit to the data for each bin in $W$ and $Q^2$.

In general, unitarity requires the presence of direct $2\pi$ production mechanisms in the $\pi^+\pi^- p$ electroproduction 
amplitudes, where the final state is created without going through the intermediate step of forming unstable hadron
states~\cite{Aitchison:1979ja,Aitchison:1978pw}. These $2\pi$ processes are beyond the aforementioned contributions from the two-body 
sub-channels and are implemented into the JM17 model. These mechanisms are incorporated by a sequence of two exchanges in the $t$- 
and/or $u$-channel by unspecified particles that belong to two Regge trajectories. The amplitudes of the $2\pi$
mechanisms are parameterized by a Lorentz-invariant contraction between the spin-tensors of the initial and final state particles, 
while two exponential propagators describe the exchanges by unspecified particles. All details on the parameterization of the $2 \pi$ 
mechanisms are available in Refs.~\cite{Mokeev:2008iw,Mokeev:2015lda}. The magnitudes of these amplitudes are fit to 
the data for each bin in $W$ and $Q^2$. 

The studies of the final state hadron angular distributions over $\alpha_i$~($i=[\pi^-p][\pi^+p']$, $[\pi^+p][\pi^-p']$, 
$[\pi^+\pi^-][p p']$) in Ref.~\cite{Mokeev:2015lda} demonstrated the need to implement relative phases for all $2\pi$ mechanisms 
included in the JM17 model determined in the data fit. The contributions from these mechanisms are maximal and substantial
($\approx$30\%) for $W < 1.5$~GeV and they decrease with increasing $W$, contributing less than 10\% for $W > 1.6$~GeV. However, 
even in this kinematic regime, these mechanisms can be seen in the $\pi^+\pi^-p$ cross sections due to an interference of the 
amplitudes with the two-body sub-channels. 

Representative examples of the description of the $\pi^+\pi^-p$ cross sections achieved within the JM17 model in several bins of 
$W < 1.6$~GeV for $Q^2$ from 3.50--4.20~GeV$^2$ are shown in Figs.~\ref{fig_jm17jm23_370146}, \ref{fig_jm17jm23_3701451}, and
\ref{fig_jm17jm23_3701456} (left). The chosen $W$ intervals are closest to the Breit-Wigner masses of the $N(1440)1/2^+$, 
$N(1520)3/2^-$, and $\Delta(1600)3/2^+$. The following discrepancies have been observed in the description of these data:
\begin{itemize}[noitemsep]
    \item The JM17 model overestimates the $\pi^-$ and $p$ polar angular distributions for forward CM angles.
    \item These discrepancies are correlated with overestimated $\pi^+$ angular distributions at backward CM polar angles.
    \item Substantial deviations between the computed and measured $d\sigma/d\alpha_{[\pi^+\pi^-][pp']}$ differential cross sections 
    become evident.
    \item  The JM17 model cannot reproduce the shape of the $\pi^+\pi^-$ invariant mass distributions for $W < 1.55$~GeV and of 
    the $\pi^-p$ invariant mass distributions at $W < 1.50$~GeV for $Q^2$ from 2.0--5.0~GeV$^2$.
\end{itemize}
Comparisons of Figs.~\ref{fig_jm17jm23_370146}, \ref{fig_jm17jm23_3701451}, and \ref{fig_jm17jm23_3701456} (left) demonstrate that these
discrepancies are related mostly to the limitations of the JM17 model for the description of the $\pi\Delta$ channels for $Q^2$
from 2.0--5.0~GeV$^2$. The phenomenological extra contact terms employed for the description of the contributions into the $\pi\Delta$
amplitudes beyond the Born terms have been further modified to achieve the quality of the data description needed for the extraction of
the electrocouplings.

These modifications are achieved by multiplying the extra contact term amplitudes $T^0_{e.c.t.\,\pi\Delta}$ employed in the JM17 model
(detailed in Ref.~\cite{Mokeev:2008iw}, Appendix II) with (a) the four factors $F_1(t'_{pp'})$, $F_2(t'_{\gamma\pi^-})$,
$F_3(t'_{\gamma\pi^+})$, and $F_4(t'_{\gamma p'})$, which allow for a better description of the CM $\theta_i$ ($i = \pi^+, \pi^-, p'$) 
angular distributions of the final state hadrons and (b) the factor $M_1(M_{\pi^+\pi^-})$ needed in order to improve the description of
$\pi^+\pi^-$ invariant mass distributions. The updated extra contact term amplitudes $T_{e.c.t.\pi\Delta}$ are parameterized as:

\begin{eqnarray}
& T_{e.c.t.\,\pi\Delta}  = T^0_{e.c.t.\,\pi\Delta}\cdot F_1(t'_{pp'})\cdot F_2(t'_{\gamma\pi^-}) & \nonumber \\ 
& ~~\cdot F_3(t'_{\gamma\pi^+})\cdot F_4(t'_{\gamma p'})\cdot M_1(M_{\pi^+\pi^-}).
\label{extracontact}
\end{eqnarray}
\noindent
Here $t'_{pp'}$, $t'_{\gamma\pi^-}$, $t'_{\gamma\pi^+}$, and $t'_{\gamma p'}$ are the squared four-momentum transfers defined by
the difference between the initial state $\gamma_v$ and $p$ and one of the final state hadrons, and their maximum values in the 
respective physics regions. They are defined by:

\begin{align}
\label{tprimes}
& t_{pp'} = (p_p-p_{p'})^2, ~~ t'_{pp'} = t_{pp'}-t_{pp'}^{max}, & \nonumber  \\
& t_{pp'}^{max} = 2m^2_p - 2E_pE_{p'} + 2\left\vert p_p \right\vert \left\vert p_{p'} \right\vert   
\end{align}

\begin{align}
\label{tgpim}
& t_{\gamma\pi^-} = (q_\gamma-p_{\pi^-})^2, ~~ t'_{\gamma \pi^-} = t_{\gamma \pi^-} - t_{\gamma \pi^-}^{max}, & \nonumber \\
& t_{\gamma \pi^-}^{max} = -Q^2 + m^2_\pi - 2E_\gamma E_{\pi^-} + 2\left\vert q_\gamma \right\vert\left\vert p_{\pi^-} \right\vert, &
\nonumber \\
& Q^2=-q_\gamma^2
\end{align}
\begin{align}
\label{tgpip}
& t_{\gamma \pi^+} = (q_\gamma - p_{\pi^+})^2, ~~ t'_{\gamma \pi^+} = t_{\gamma \pi^+} - t_{\gamma \pi^+}^{max}, & \nonumber \\
& t_{\gamma \pi^+}^{max} = -Q^2 + m^2_\pi - 2E_\gamma E_{\pi^+} + 2\left\vert q_\gamma \right\vert \left\vert p_{\pi^+} \right\vert   
\end{align}
\begin{align}
\label{tgpr}
& t_{\gamma p'} = (q_\gamma - p_{p'})^2, ~~ t'_{\gamma p'} = t_{\gamma p'} - t_{\gamma p'}^{max}, & \nonumber \\
& t_{\gamma p'}^{max} = -Q^2 + m^2_p - 2E_\gamma E_{p'} + 2\left\vert q_\gamma \right\vert\left\vert p_{p'} \right\vert.   
\end{align}

\noindent
Here $q_\gamma$, $p_{\pi^+}$, $p_{\pi^-}$, and $p_{p'}$ are the four-momenta of the initial state photon and the final state $\pi^+$,
$\pi^-$, and $p$, respectively, and $p_p$ is the four-momentum of the initial state proton. $q^2$, $E_p$, and $m_p$ represent the 
virtual photon four-momentum squared, the initial state proton CM energy, and the proton mass. $E_{\pi^+}$, $E_{\pi^-}$, $E_{p'}$, and
$\left\vert p_{\pi^-} \right\vert$, $\left\vert p_{\pi^+} \right\vert$, $\left\vert p_{p'} \right\vert$ are the CM energies and absolute
values of the three-momenta for the final state $\pi^+$, $\pi^-$, and $p$, respectively, while $\left\vert q_\gamma \right\vert$ is the
absolute value of the virtual photon three-momentum in the CM frame.

The factors $F_j$  in Eq.(\ref{extracontact}) are parameterized as:
\begin{equation}
F_j= \begin{cases}-t'_j, & \mbox{if}\,\,\, t_j > \Lambda_j(W,Q^2)\\ 
\Lambda_j(W)-t_j^{max} &\mbox{if}\,\,\, t_j < \Lambda_j(W,Q^2), \end{cases}
\label{ffactor}
\end{equation}
where $j$=$pp'$, $\gamma\pi^-$, $\gamma\pi^+$, $\gamma p'$ and the parameters $\Lambda_j(W,Q^2)$ are adjusted to reproduce the data on 
the CM $\pi^+$, $\pi^-$, and $p$ angular distributions in each bin of $W$ and $Q^2$ independently.

\begin{figure*}[htbp]
\begin{center}
\includegraphics[width=13.4cm]{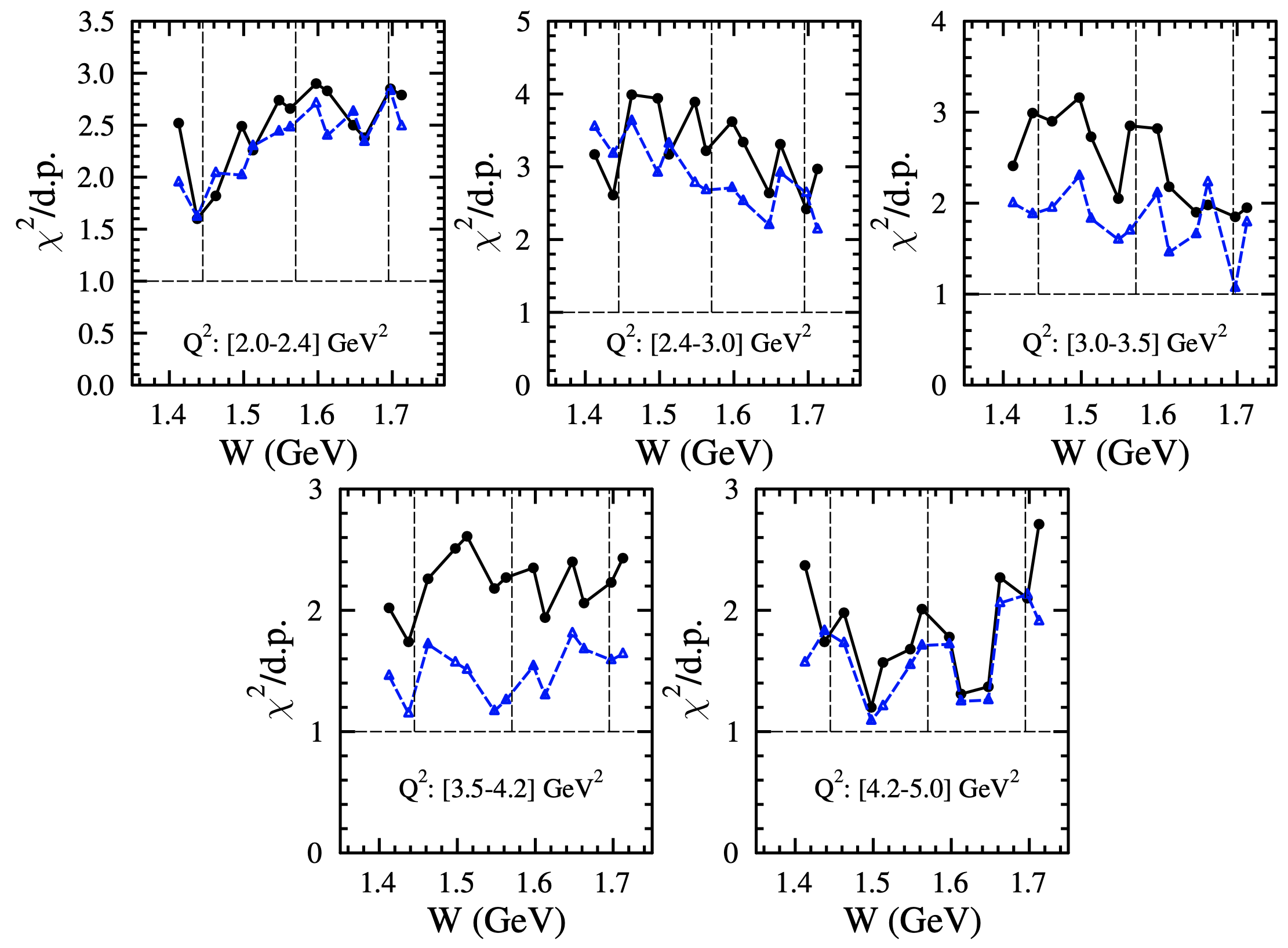}
\caption{Description of the nine one-fold differential $\pi^+\pi^-p$ electroproduction cross sections in terms of $\chi^2/d.p.$
determined from the point-by-point comparison between the experimental data (only statistical uncertainties taken into account) and 
the computed values within the JM23 model as a function of $W$ for $Q^2$ as indicated on the plots. $\chi^2/d.p.$ with/without
implementation of the $\Delta(1600)3/2^+$ is shown by the blue dashed/black solid lines, respectively. The Breit-Wigner mass of the
$\Delta(1600)3/2^+$ is indicated by the middle vertical line and the outer vertical lines show the total decay width.}  
\label{chi2_jm23}
\end{center}
\end{figure*}

The $F_i$ factors make the extra contact term contributions equal to zero at the maximum accessible values of the respective squared 
four-momentum transfers. They cause the amplitudes to increase with $-t'_j$ (or with absolute $t'_j$ values) in the range of 
$t'_j > \Lambda_j$, making the final state hadron CM angular distributions closer to those measured with CLAS.

The factor $M_1(M_{\pi^+\pi^-})$ in Eq.(\ref{extracontact}) is essential to improve the description of the $\pi^+\pi^-$ 
invariant mass distributions. It is given by:

\begin{widetext}
\begin{equation}
M_1(M_{\pi^+\pi^-})= \begin{cases} \frac{M^2_{\pi^+\pi^-} - a_{\pi^+\pi^-}(W,Q^2)m^2_\pi}{(W-b_{\pi^+\pi^-}(W,Q^2)M_p)^2 -
M^2_{\pi^+\pi^-}} 
& \mbox{if}\,\,\, W < 1.6\,\, \mbox{GeV}\\ 
 1 &\mbox{if}\,\,\, W>1.6\,\, \mbox{GeV,} \end{cases} \label{m1factor}
\end{equation}
\end{widetext}
where the parameters $a_{\pi^+\pi^-}(W,Q^2)$ ($a_{\pi^+\pi^-}>4.0$) and $b_{\pi^+\pi^-}(W,Q^2)$ ($0.0<b_{\pi^+\pi^-}<1.0$) are 
adjusted to the data in each bin of $W$ and $Q^2$ independently. At values of $b_{\pi^+\pi^-}(W,Q^2)$ equal to unity, the factor
$M_1(M_{\pi^+\pi^-})$ develops a pole at the maximum kinematically allowed invariant masses $M_{\pi^+\pi^-}$. Hence, the denominator 
in Eq.(\ref{m1factor}) defines the shape of the $M_{\pi^+\pi^-}$ invariant mass distributions at their largest kinematically allowed
values, while the numerator in Eq.(\ref{m1factor}) regulates the slope of the $M_{\pi^+\pi^-}$ mass distributions.

The modifications of the JM17 model extra contact term described above and the implementation of the $\Delta(1600)3/2^+$ led to the 
updated model version referred to as JM23. With the JM23 model, a reasonable description of the experimental data on the nine 
independent one-fold differential $\pi^+\pi^-p$ electroproduction cross sections has been achieved in the extended $Q^2$ range 
as exemplified in Figs.~\ref{fig_jm17jm23_370146}, \ref{fig_jm17jm23_3701451}, and \ref{fig_jm17jm23_3701456} (right). The 
$\chi^2/d.p.$ ($d.p.$ = data point) values computed within the range of $W < 1.7$~GeV for $Q^2$ from 2.0--5.0~GeV$^2$ from the 
point-by-point comparison between the measured $\pi^+\pi^-p$ differential cross sections and those evaluated within JM23 are shown in
Fig.~\ref{chi2_jm23} as a function of $W$ in each $Q^2$ interval covered by the analyzed CLAS data. Only the statistical data
uncertainties were included in the evaluation of $\chi^2/d.p.$ to enhance the sensitivity of these quantities to the parameterization 
of non-resonant contributions. In each bin the $\chi^2/d.p.$ values are comparable with those achieved in previous analyses of 
$\pi^+\pi^-p$ electroproduction data~\cite{Mokeev:2012vsa, Mokeev:2015lda} where the electrocouplings were deduced from the data 
fits within the previous versions of the JM model. From 6 to 8 parameters of the non-resonant amplitudes were varied in the extraction 
of the $\gamma_vpN^*$ electrocouplings, which were fit to 102 data points in each bin of $(W,Q^2)$.

We also explored the impact of the $\Delta(1600)3/2^+$ on the description of the $\pi^+\pi^-p$ differential cross sections. The 
four-star status for this state was assigned after 2017, hence, it was not included in the JM17 model. The $\Delta(1600)3/2^+$ was
incorporated into the JM23 version with starting hadronic decay parameters taken from the PDG~\cite{ParticleDataGroup:2022pth} and 
with the initial $\gamma_vpN^*$ electrocouplings from the CSM predictions~\cite{Lu:2019bjs}. The results
in Fig.~\ref{chi2_jm23} demonstrate improvements in the description of the $\pi^+\pi^-p$ differential cross sections after 
implementation of this state. The improvements become more pronounced with increasing $Q^2$, suggesting a relative increase of the 
signal from the $\Delta(1600)3/2^+$ with $Q^2$. For the highest $Q^2$ bin, because of the increase of the data uncertainties, the
improvement of the data description is less pronounced. It is also worth noting that the implementation of the $\Delta(1600)3/2^+$ 
into JM23 has allowed for the reduction of the magnitudes of the direct $2 \pi$ production amplitudes by 20--50\% and for the extra
contact term amplitudes in the $\pi\Delta$ sub-channels by 10--30\%, which represent the contributions described within the entirely
phenomenological parameterization. These observations have confirmed the impact of the $\Delta(1600)3/2^+$ contribution in
$\pi^+\pi^-p$ electroproduction. 

\begin{table}
\begin{center}
\begin{tabular}{|c|c|} \hline
     Resonance            & $W$ Interval, GeV \\ \hline
                          & 1.41-1.51    \\
      $N(1440)1/2^+$      & 1.46-1.56  \\ 
                          & 1.51-1.61  \\ \hline
                          & 1.41-1.51 \\ 
      $N(1520)3/2^-$      & 1.46-1.56   \\
                          & 1.51-1.61 \\ \hline 
                          & 1.46-1.56   \\
      $\Delta(1600)3/2^+$ & 1.51-1.61  \\  
                          & 1.56-1.66   \\ \hline
\end{tabular}
\caption{$W$ intervals where the nine independent one-fold  differential $\pi^+\pi^-p$ electroproduction cross sections were fit
independently for $Q^2$ from 2.0--5.0~GeV$^2$ to extract the $N(1440)1/2^+$, $N(1520)3/2^-$, and $\Delta(1600)3/2^+$
electrocouplings and their $BF$ for decays into the $\pi\Delta$ and $\rho p$ intermediate states with JM23.}
\label{W_ranges} 
\end{center}
\end{table}

\section{Resonance Parameter Extraction From Cross Section Fits}
\label{fit_strategy}

The electrocouplings of the $N(1440)1/2^+$, $N(1520)3/2^-$, and $\Delta(1600)3/2^+$, as well as their branching fractions for decays 
into $\pi\Delta$ and $\rho p$, were extracted from fits of the nine independent one-fold differential $\pi^+\pi^-p$ cross sections as 
described in Section~\ref{kinematics}. A good description of the $\pi^+\pi^-p$ electroproduction  data achieved with 
the JM23 model at $W< 1.7$~GeV for $Q^2$ from 2.0--5.0~GeV$^2$ allows for the determination of the resonance parameters from the data 
fit  for $W$ from 1.40--1.66~GeV and $Q^2$ from 2.0--5.0~GeV$^2$. The electrocouplings and hadronic decay parameters were obtained 
from independent fits within the three overlapping $W$ intervals given in Table~\ref{W_ranges}. The non-resonant contributions in 
these $W$ intervals are different, while the electrocouplings determined from the data fits should be the same within their 
uncertainties, validating  extraction of these quantities.

In the data fit we simultaneously varied the electrocouplings, the resonance partial decay widths into $\pi\Delta$ and $\rho p$, and 
the Breit-Wigner masses for all $N^*$ states listed in Table~\ref{nstlist}. The starting values for the resonance decay amplitudes 
into $\pi\Delta$ and $\rho p$ of orbital angular momentum $L$ and total spin $S$, are defined by
\begin{equation}
 \sqrt{\Gamma^i_{LS}} = \sqrt{\Gamma_{tot}\cdot BF^i_{LS}}, 
 \label{hadr_decay}
\end{equation}
where the resonance total decay widths $\Gamma_{tot}$ were taken from Ref.~\cite{ParticleDataGroup:2022pth}, and the branching 
fractions $BF^i_{LS}$ ($i = \pi\Delta, \rho p$) for the $N(1440)1/2^+$ and $N(1520)3/2^-$ were taken from the previous studies of 
CLAS $\pi N$ and $\pi^+\pi^-p$ electroproduction data~\cite{Aznauryan:2009mx,Mokeev:2012vsa,Mokeev:2015lda}. For other excited 
states, the outcome of the analyses of Refs.~\cite{Manley:1995ys,Manley:1992yb} were used for the $BF^i_{LS}$ starting values. For 
each resonance, the total decay width was computed as the sum of all partial decay widths. The floating of the resonance masses and 
total decay widths $\Gamma_{tot}$ caused by variation of the partial hadronic decay widths into $\pi\Delta$ and $\rho p$ were limited 
by the intervals given in Ref.~\cite{ParticleDataGroup:2022pth}. In this way, we imposed restrictions for the variation of the $N^*$
partial hadronic decay widths $\Gamma^i_{LS}$ ($i = \pi\Delta, \rho p$)into $\pi \Delta$ and $\rho p$ decomposed over the $LS$-partial
waves.

The starting values for the $N(1440)1/2^+$ and $N(1520)3/2^-$ electrocouplings used in the fit of the $\pi^+\pi^-p$ electroproduction 
cross sections were taken from the analysis of $\pi N$ electroproduction~\cite{Aznauryan:2009mx}. The transverse electrocouplings 
$A_{1/2}$ and $A_{3/2}$ of these states were varied by employing normal distributions with the $\sigma$ parameters equal to 30\% 
of their starting values. There were no restrictions on the minimum or maximum trial electrocoupling values, allowing us to explore 
the area of $\approx \pm 3\sigma$ around the starting values. The longitudinal $S_{1/2}$ electrocouplings of smaller absolute values 
were varied within a broader range so that their absolute values overlapped with the absolute values of the transverse electrocouplings.

The starting values for the electrocouplings of the $\Delta(1600)3/2^+$ were based on the CSM predictions~\cite{Lu:2019bjs}. Currently, 
CSM takes into account only the contributions from the quark core that gradually dominate as $Q^2$ increases, typically for 
$Q^2 \gtrsim 2$~GeV$^2$, which is the range covered in this analysis. Therefore, the actual starting values of the $\Delta(1600)3/2^+$
electrocouplings were taken from the values of Ref.~\cite{Lu:2019bjs} multiplied by a common factor of 0.6 applied over the entire range 
of $Q^2$ for all three electrocouplings $A_{1/2}$, $A_{3/2}$, and $S_{1/2}$. This factor accounts for the fact that the wave function of 
the $\Delta(1600)3/2^+$ represents a superposition of the contributions from the quark core and meson-baryon cloud. In the data fit, 
implementation of this factor was needed to reproduce the results on the final state proton CM angular distributions in the forward 
hemisphere. The $\Delta(1600)3/2^+$ electrocouplings were varied employing normal distributions with the $\sigma$ parameters equal to 
50\% of their starting values. 

In the fits we simultaneously varied the electrocouplings of all $N^*$s of four-star status (other than the $N(1440)1/2^+$,
$N(1520)3/2^-$, and $\Delta(1600)3/2^+$) as described above within the mass range below 1.7~GeV. This variation employed a normal
distribution with the $\sigma$ parameters equal to 20\% of their starting values, which were taken from the analyses of the CLAS 
exclusive meson electroproduction data~\cite{Mokeev:2022xfo} with the numerical results available in Ref.~\cite{HillerBlin:2019jgp}. 

\begin{figure*}[htbp]
\begin{center}
\includegraphics[width=8.0cm]{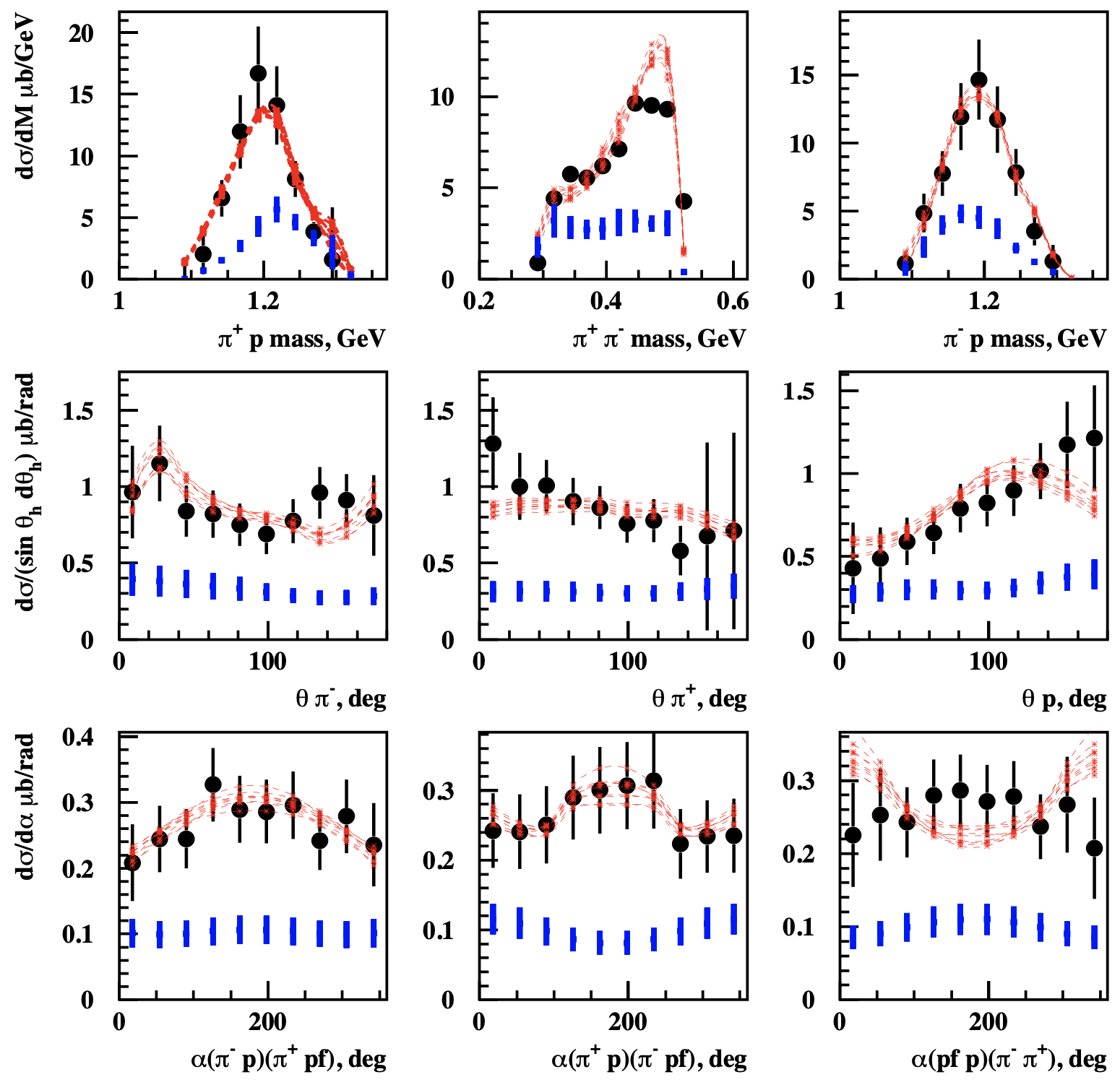}
\hspace{5mm}
\includegraphics[width=8.0cm]{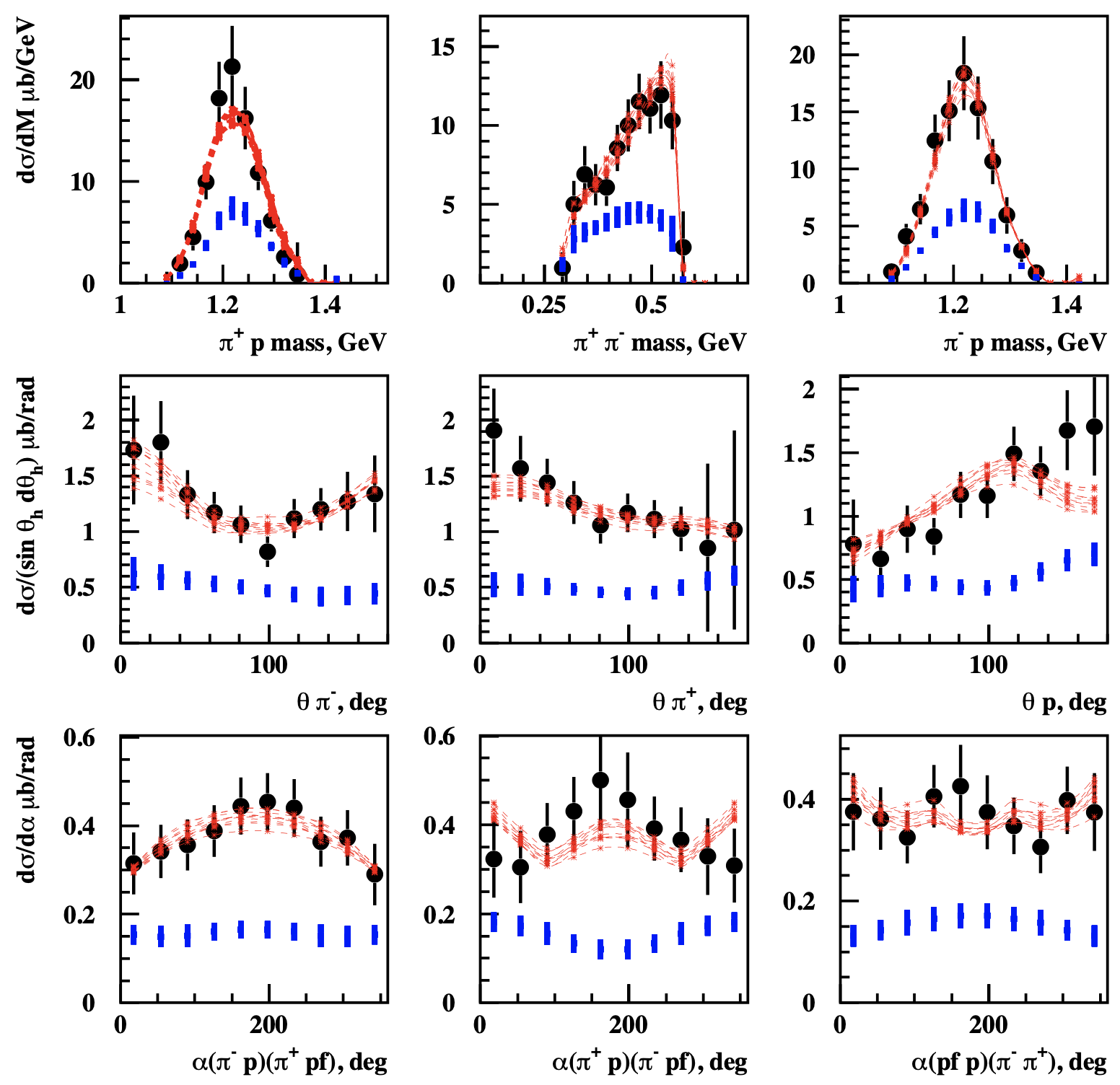}

\vspace{2mm}
\includegraphics[width=8.0cm]{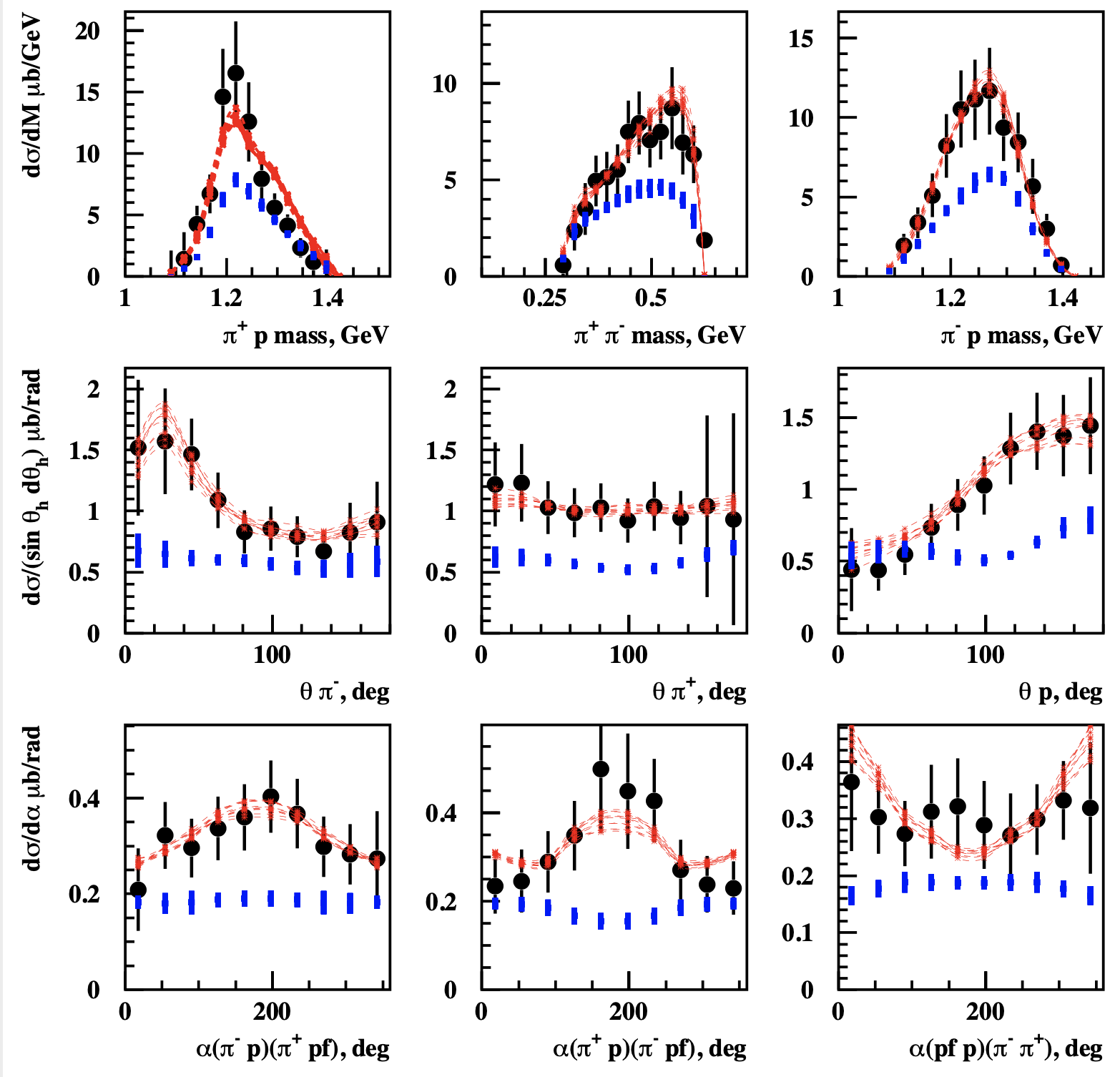}
\vspace{-12mm}
\caption{Fit of the nine one-fold differential $\pi^+\pi^-p$ electroproduction cross sections measured with CLAS
\cite{Trivedi:2018rgo} (in black) achieved within the JM23 model for $W$ from 1.450--1.475 GeV (top left), 1.500--1.525~GeV, (top 
right) and 1.550--1.575 GeV (bottom) for $Q^2$ from 3.0-3.5~GeV$^2$. The data point uncertainties are evaluated as a quadratic sum of
statistical and relevant systematic uncertainties. The groups of red curves represent the JM23 fits closest to the data. The 
resonant contributions are shown in blue.} 
\label{full_nstar_146151156_31}
\end{center}
\end{figure*}

In the data fit we also varied the following parameters of the non-resonant mechanisms employed in the JM23 model:
\begin{itemize}
\item the magnitudes of the additional contact-term amplitudes in the $\pi^- \Delta^{++}$ and $\pi^+ \Delta^0$ channels (1 parameter 
per $Q^2$-bin);
\item the magnitudes of the $\pi^+ N^0(1520)3/2^-$ channel (1 parameter per $Q^2$-bin);
\item the magnitudes of all direct $2 \pi$ production amplitudes (up to 6 parameters per $Q^2$-bin).
\end{itemize}
The starting values for these parameters were determined in their initial adjustment to the nine independent one-fold $\pi^+\pi^-p$
differential cross sections. We applied $W$-independent multiplicative factors to the magnitudes of the non-resonant amplitudes 
listed above. They remained the same in the entire $W$ interval covered by the fit within any $Q^2$-bin, but they depended on $Q^2$ 
and were fit to the data in each $Q^2$-bin independently. The multiplicative factors were varied around unity, employing normal
distributions with $\sigma$ values in the range of 20\%. In this way, we retained a smooth $W$-dependence of the non-resonant 
contributions established in the adjustment to the data and explored the possibility of improving the data description in the 
simultaneous variation of the resonant and non-resonant parameters.

The special data fit procedure described in Ref.~\cite{Mokeev:2012vsa} was employed for the extraction of the resonance parameters. It
allowed us to obtain not only the best fit but also to establish bands of the computed cross sections that were compatible with the 
data within their uncertainties. For each trial set of JM23 resonant and non-resonant parameters, we computed the nine one-fold 
differential $\pi^+\pi^-p$ cross sections and $\chi^2$/$d.p.$ values. The latter were estimated in point-by-point comparisons between 
the measured and computed cross sections in all bins of $W$ from 1.41--1.66~GeV for $Q^2$ from 2.0--5.0~GeV$^2$ covered by the CLAS 
data. The data uncertainties account for both the statistical and that part of the systematic uncertainty dependent on the final 
state hadron kinematics, which were added in quadrature. In the fit, we selected the computed one-fold differential cross sections
closest to the data with $\chi^2/d.p.$ less than a predetermined maximum value. These values of  $\chi^2_{max}/d.p.$ were obtained 
by requiring that the computed cross sections with smaller $\chi^2/d.p.$ be within the data uncertainties for the majority of the 
data points. In this fit procedure, we obtained the $\chi^2/d.p.$ intervals within which the computed cross sections described the 
data equally well within the data uncertainties. The ranges of $\chi^2/d.p.$ listed in Table~\ref{chidp} demonstrate that the cross
sections selected in the data fit are indeed distributed within the data uncertainties over the entire area of $(W,Q^2)$ covered in 
the data analysis. 

\begin{table}
\begin{center}
\begin{tabular}{|c|c|c|c|c|} \hline
$W$ Interval, &        &            &       &         \\
 GeV          & 1.41-1.51 & 1.46-1.56 & 1.51-1.61 & 1.56-1.66 \\ \hline
$\chi^2/d.p.$ &          &           &        &         \\
Ranges        & 0.51-0.57 & 0.52-0.67 & 0.52-0.69 & 0.69-0.76 \\ \hline 
\end{tabular}
\caption{The ranges of $\chi^2/d.p.$ for the nine one-fold differential $\pi^+\pi^-p$ electroproduction cross sections selected in
the data fit computed within JM23 in overlapping $W$ intervals for $Q^2$ from 2.0--5.0~GeV$^2$. The uncertainties for the measured
data are given by the quadratic sum of the statistical and that part of the systematic uncertainty dependent on the final state 
hadron kinematics.}
\label{chidp} 
\end{center}
\end{table}

Representative examples for the fit quality of the cross sections within the $W$ intervals closest to the Breit-Wigner masses of the
$N(1440)1/2^+$, $N(1520)3/2^-$, and $\Delta(1600)3/2^+$ for $Q^2$ from 3.0--3.5~GeV$^2$ are shown in Fig.~\ref{full_nstar_146151156_31}. 
The computed cross sections selected in the data fit are shown by the red curves. The resonant contributions were obtained 
from the differential cross sections computed from only the resonance amplitudes and are shown by the blue bars. The deduced 
uncertainties for the resonance contributions are comparable both with the uncertainties of the measured differential cross sections 
and with the spread of the fits computed within the JM23 model. Pronounced differences are evident in the shapes of the computed
differential cross sections selected in the data fit and the respective resonant contributions, in particular, for all angular 
distributions. The shapes of the resonant contributions are different in each of the nine one-fold differential cross sections but 
they are highly correlated by the reaction dynamics that underlie the resonance excitations in the $s$-channel and their subsequent 
decays into either the $\pi\Delta$ or $\rho p$ intermediate states. The resonant/non-resonant contribution differences allow for 
isolation of the resonant contributions. The electrocouplings and decay widths into $\pi\Delta$ and $\rho p$ have been determined 
from the resonant contributions by fitting them within the unitarized Breit-Wigner ansatz~\cite{Mokeev:2012vsa, Aitchison:1972ay}, 
taking into account the constraints imposed on the resonant amplitudes by a general unitarity condition. 
 
\section{$N(1440)1/2^+$, $N(1520)3/2^-$, and $\Delta(1600)3/2^+$ Electrocouplings and Hadronic Decay Widths to $\pi\Delta$ and $\rho p$}
\label{elcoupl_hadrdec}

The resonance parameters determined from the data fit include the electrocouplings, the partial decay widths into $\pi\Delta$ and 
$\rho p$, and the total resonance decay widths. They are averaged from the group of fits selected by the $\chi^2/d.p.$ limits for 
each bin and their mean values are taken as the resonance parameters extracted from the data. The r.m.s dispersions in these parameters
are taken as the uncertainties. The electrocoupling uncertainties obtained in this manner take into account both the statistical and
systematic uncertainties in the data, as well as the systematic uncertainties associated with the JM23 model. Furthermore, we
consistently account for the correlations between the variations of the resonant and non-resonant contributions when extracting the 
resonance parameters. In the cases where the ranges of the extracted electrocouplings covered more than 90\% of the intervals for the 
electrocoupling variation (starting values $\pm \sigma$) employed in the data fit, we further increased the ranges of the variation 
and repeated the data fit, so that eventually the electrocouplings extracted from the data were inside the intervals of the variations 
employed in the data fit. In this way, we made sure that the employed ranges were sufficient to determine both the mean values of the
resonance parameters and their uncertainties. 

\begin{figure}[htbp]
\begin{center}
\includegraphics[width=7.0cm]{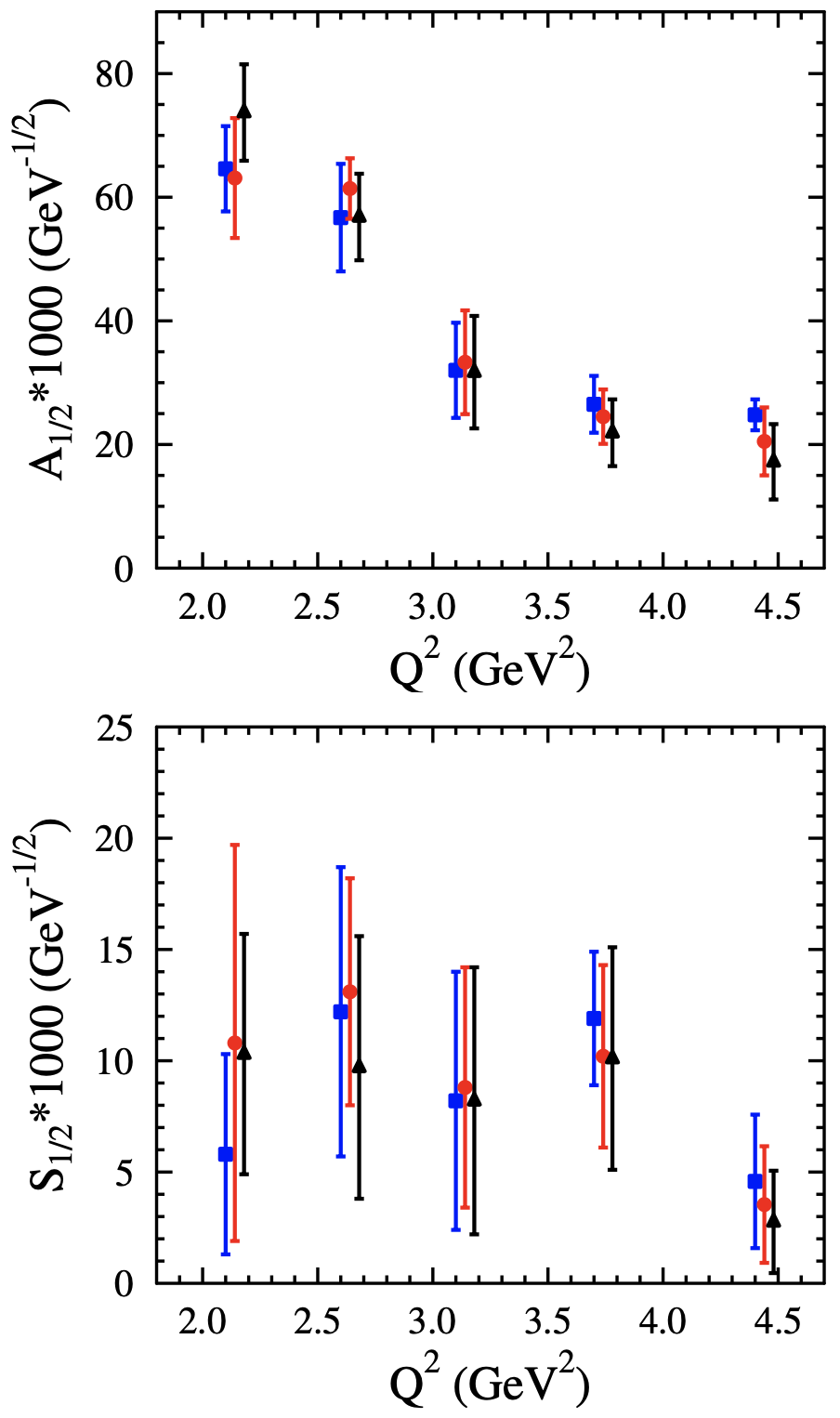}
\caption{Electrocouplings of the $N(1440)1/2^+$ determined from independent fits of the $\pi^+\pi^-p$ electroproduction cross sections 
in three $W$ intervals, 1.41--1.51~GeV (blue squares), 1.46--1.56~GeV (red circles), and 1.51--1.61~GeV (black triangles), for $Q^2$ 
from 2.0--5.0~GeV$^2$ within the JM23 model.} 
\label{n1440_elcoupl141151146156151161}
\end{center}
\end{figure}

\begin{figure}[h!]
\begin{center}
\includegraphics[width=7.5cm]{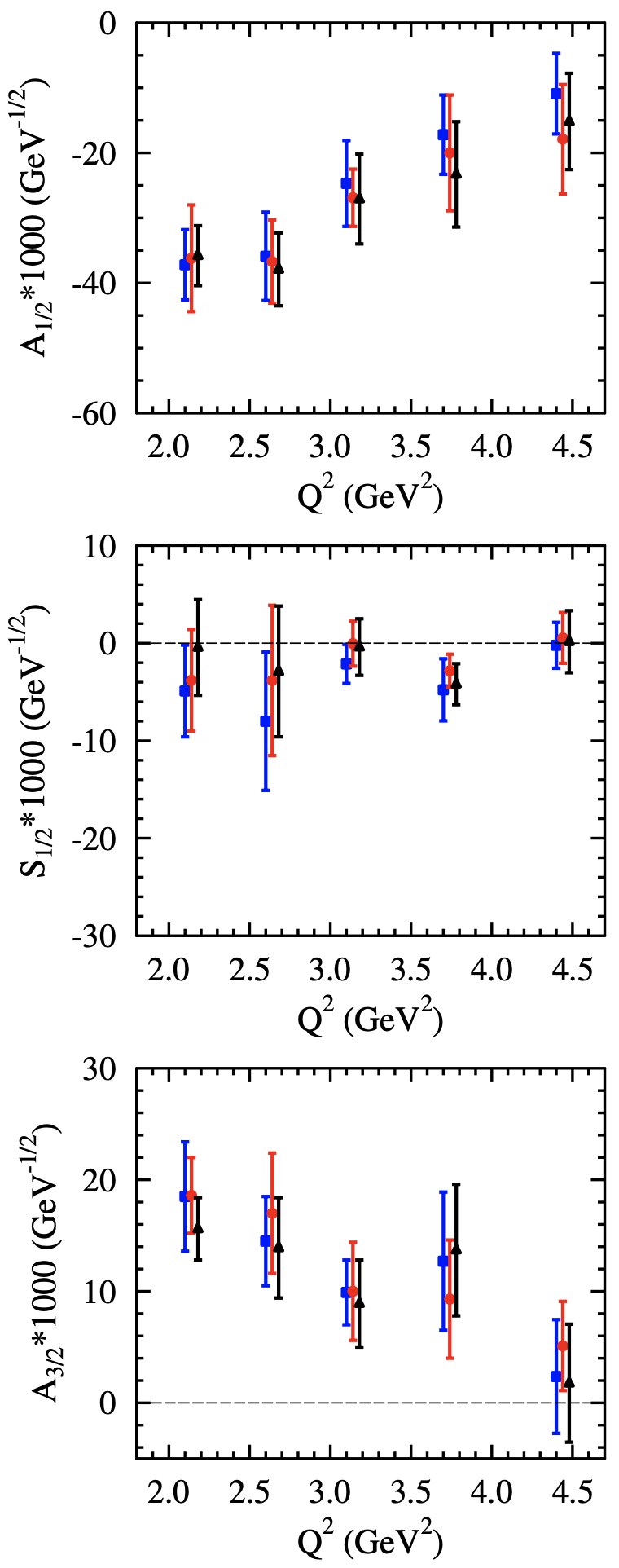}
\caption{Electrocouplings of the $N(1520)3/2^-$ determined from independent fits of the $\pi^+\pi^-p$ electroproduction cross sections 
in three $W$ intervals, 1.41--1.51~GeV (blue squares), 1.46--1.56~GeV (red circles), and 1.51--1.61~GeV (black triangles), for $Q^2$ 
from 2.0--5.0~GeV$^2$ within the JM23 model.} 
\label{n1520_elcoupl141151146156151161}
\end{center}
\end{figure}

\begin{table}
\begin{center}
\begin{tabular}{|c|c|c|c|c|c|c|} \hline
$Q^2$ Inter- & Mass, &$\Gamma_{tot}$,&$\Gamma_{\pi\Delta},$ &$BF_{\pi\Delta}$,&$\Gamma_{\rho p}$,&$BF_{\rho p}$,         \\
val, GeV$^2$         &  GeV  & MeV           &MeV                  & \%               &MeV               & \% \\ \hline
0.25-0.60          & 1.458$\pm$0.012 &363$\pm$39&    142$\pm$48     & 23-58   &  6$\pm$4 & $<$2 \\ \hline
0.5-1.5          & 1.450$\pm$0.011 &352$\pm$37&    120$\pm$41     & 20-52   &  5$\pm$2 & $<$2 \\ \hline
2.0-3.5          & 1.457$\pm$0.008 &331$\pm$54&    129$\pm$52     & 20-65   &  6$\pm$2 & 1.1-2.6 \\ \hline
3.0-5.0          & 1.446$\pm$0.013 &352$\pm$33&    151$\pm$32     & 31-57   &  5$\pm$1 & 1.2-2.0 \\ \hline
\end{tabular}
\caption{Masses and total/partial hadronic decay widths of the $N(1440)1/2^+$ into $\pi\Delta$ and $\rho p$ determined from fits of 
the $\pi^+\pi^-p$ electroproduction cross sections carried out independently within different $Q^2$ intervals. The new results from
this work are given in the last two rows. The results in the upper rows are available from previous studies
\cite{Mokeev:2012vsa,Mokeev:2015lda} of the $\pi^+\pi^-p$ electroproduction cross sections.}
\label{hadr_n1440} 
\end{center}
\end{table}

\begin{table}
\begin{center}
\begin{tabular}{|c|c|c|c|c|c|c|} \hline
$Q^2$ Inter- & Mass, &$\Gamma_{tot}$,&$\Gamma_{\pi\Delta},$ &$BF_{\pi\Delta}$,&$\Gamma_{\rho p}$,&$BF_{\rho p}$,         \\
val, GeV$^2$         &  GeV  & MeV           &MeV                  & \%               &MeV               & \% \\ \hline
0.25-0.60        & 1.521$\pm$0.004 &127$\pm$4&    35$\pm$4     & 24-32   &  16$\pm$5 & 8-17 \\ \hline
0.5-1.5          & 1.520$\pm$0.001 &125$\pm$4&    36$\pm$5     & 25-34   &  13$\pm$6 & 5-16 \\ \hline
2.0-3.5          & 1.518$\pm$0.003 &122$\pm$7&    29$\pm$5     & 19-30   &  15$\pm$6 & 7-18 \\ \hline
3.0-5.0          & 1.522$\pm$0.003 &121$\pm$7&    30$\pm$5     & 20-30   &  13$\pm$5 & 6-16 \\ \hline
\end{tabular}
\caption{Masses and total/partial hadronic decay widths of the $N(1520)3/2^-$ into $\pi\Delta$ and $\rho p$ determined from fits of 
the $\pi^+\pi^-p$ electroproduction cross sections carried out independently within different $Q^2$ intervals. The new results from
this work are given in the last two rows. The results in the upper rows are available from previous studies
\cite{Mokeev:2012vsa,Mokeev:2015lda} of the $\pi^+\pi^-p$ electroproduction cross sections.}
\label{hadr_n1520} 
\end{center}
\end{table}

\subsection{Parameters for the $N(1440)1/2^+$ and $N(1520)3/2^-$}
\label{n1440n1520}

\begin{figure*}[htbp]
\begin{center}
\includegraphics[width=15.0cm]{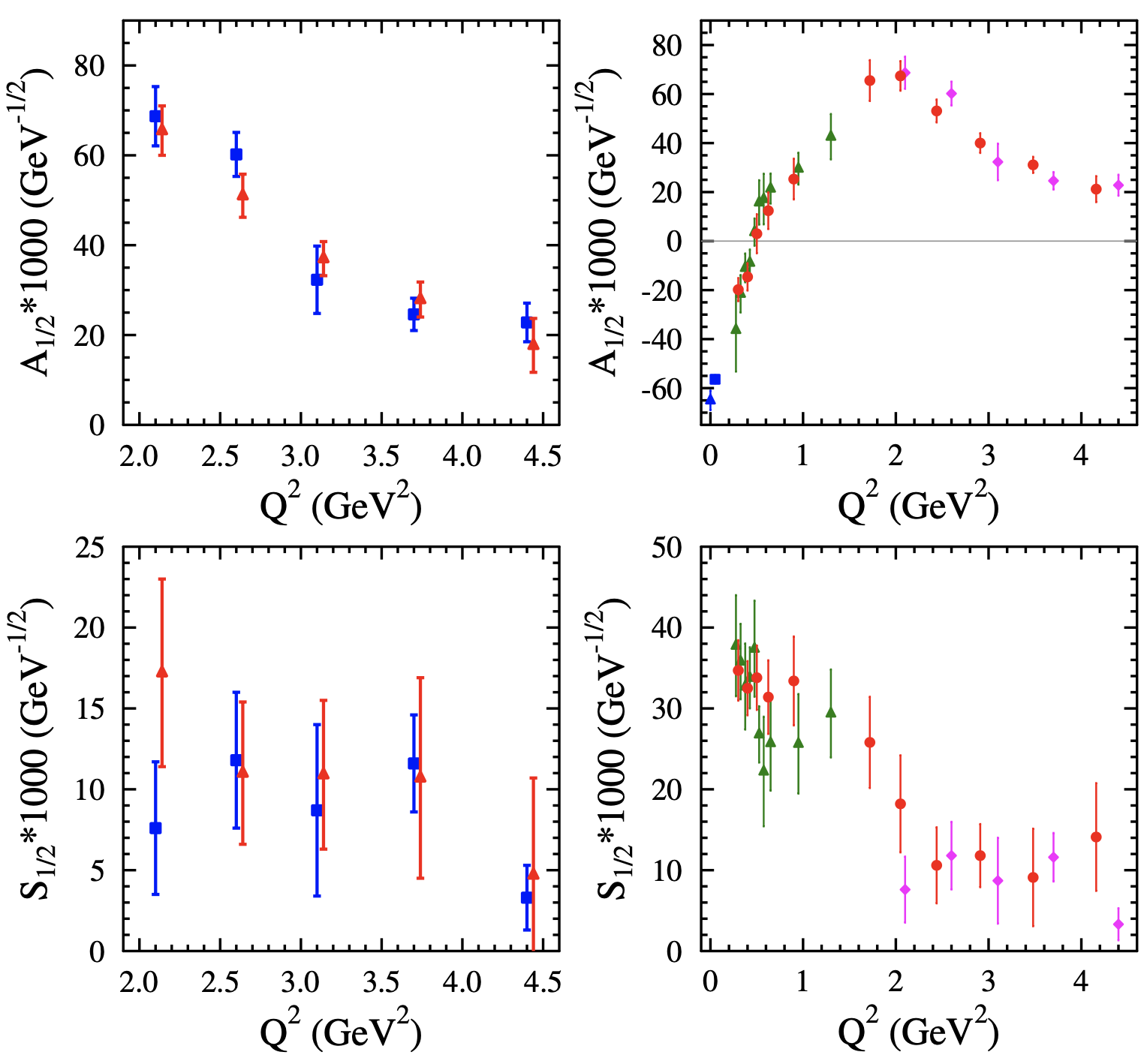}
\vspace{-2mm}
\caption{(Left) $N(1440)1/2^+$ electrocouplings determined from the $\pi N$ differential cross sections, beam, target, and beam-target
asymmetries~\cite{Aznauryan:2009mx} (red triangles) and from the $\pi^+\pi^-p$ differential cross sections (blue squares) for $Q^2$ 
from 2.0--5.0~GeV$^2$ presented in this work. The electrocouplings from the $\pi N$ data after interpolation over $Q^2$ are compared 
with the results from the $\pi^+\pi^-p$ data. (Right) $N(1440)1/2^+$ electrocouplings from the $\pi N$ and $\pi^+\pi^-p$ data for $Q^2$
from 0.25--5.0~GeV$^2$. The results from $\pi N$ electroproduction \cite{Aznauryan:2009mx} are shown by the red circles. The
electrocouplings from the $\pi^+\pi^-p$ differential cross sections measured with CLAS for $Q^2$ from 0.25--1.5~GeV$^2$
\cite{Mokeev:2012vsa,Mokeev:2015lda} are shown by the green triangles. The electrocouplings determined within the JM23 model are shown 
by the magenta diamonds. The photocouplings from the PDG~\cite{ParticleDataGroup:2022pth} and from the CLAS $\pi N$ photoproduction
data~\cite{CLAS:2009tyz} are shown by the blue triangle and square, respectively.}
\label{n1440el_npi_pipimp}
\end{center}
\end{figure*}

The hadronic decay widths of the $N(1440)1/2^+$ and $N(1520)3/2^-$ were deduced from the fits of the $\pi^+\pi^-p$ differential cross
sections using the procedures described in Section~\ref{fit_strategy}. The results are presented in Table~\ref{hadr_n1440} for
the $N(1440)1/2^+$ and in Table~\ref{hadr_n1520} for the $N(1520)3/2^-$. For both resonances, their masses and total/partial hadronic 
decay widths into $\pi \Delta$ and $\rho p$ are self-consistent in the four $Q^2$ intervals covered by the CLAS $\pi^+\pi^-p$
electroproduction cross sections from the previous studies~\cite{Mokeev:2012vsa,Mokeev:2015lda} and those reported in this work. The
successful fit of the data achieved within a broad range of $Q^2$ from 0.25--5.0~GeV$^2$ with $Q^2$-independent resonance masses and
total/partial hadronic decay widths given the pronounced evolution of the non-resonant mechanisms with $Q^2$, demonstrates that 
both the $N(1440)1/2^+$ and $N(1520)3/2^-$ are excited states of the proton produced in the $s$-channel for the $\gamma_v p$ 
interaction.

The electrocouplings for the $N(1440)1/2^+$ and $N(1520)3/2^-$ determined from the fits of the $\pi^+\pi^-p$ cross sections carried 
out independently in three overlapping $W$ intervals for $Q^2$ from 2.0--5.0~GeV$^2$ are shown in 
Figs.~\ref{n1440_elcoupl141151146156151161} and \ref{n1520_elcoupl141151146156151161}, respectively. The non-resonant contributions in 
these three $W$ intervals are different, however the extracted electrocouplings are consistent within the uncertainties, 
suggesting credible extraction of these quantities.

\begin{figure*}[htbp]
\begin{center}
\includegraphics[width=15.0cm]{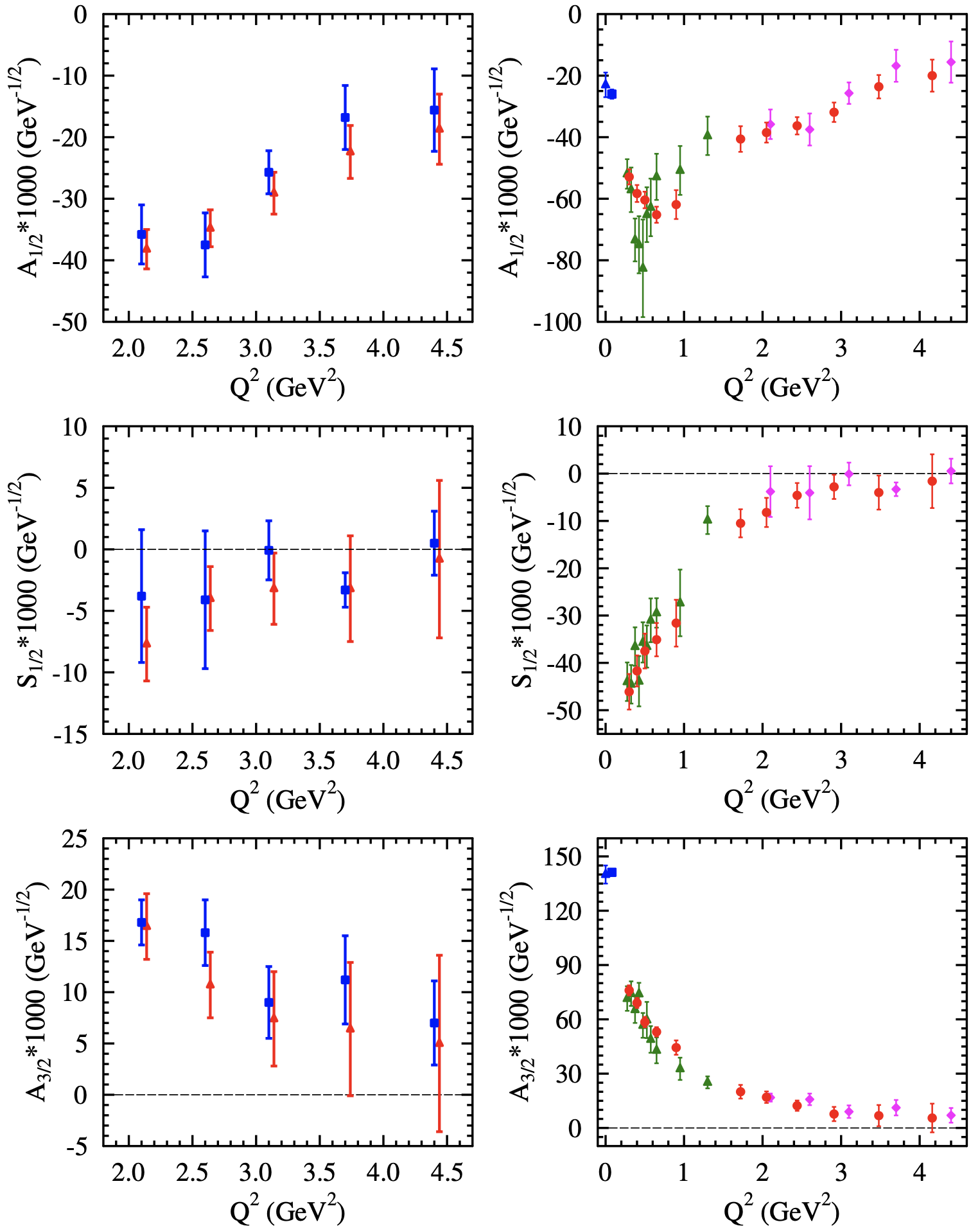}
\caption{(Left) $N(1520)3/2^-$ electrocouplings determined from the $\pi N$ differential cross sections, beam, target, and beam-target 
asymmetries~\cite{Aznauryan:2009mx} (red triangles) and from the $\pi^+\pi^-p$ differential cross sections (blue squares) for $Q^2$ 
from 2.0--5.0~GeV$^2$ presented in this work. The electrocouplings from the $\pi N$ data after interpolation over $Q^2$ are compared 
with the results from the $\pi^+\pi^-p$ data. (Right) $N(1520)3/2^-$ electrocouplings from the $\pi N$ and $\pi^+\pi^-p$ data for $Q^2$ 
from 0.25--5.0~GeV$^2$. The results from $\pi N$ electroproduction~\cite{Aznauryan:2009mx} are shown by the red circles. The 
electrocouplings from the $\pi^+\pi^-p$ differential cross sections measured with CLAS for $Q^2$ from 0.25--1.5~GeV$^2$
\cite{Mokeev:2012vsa,Mokeev:2015lda} are shown by the green triangles. The electrocouplings determined within the JM23 model are shown 
by the magenta diamonds. The photocouplings from the PDG~\cite{ParticleDataGroup:2022pth} and from the CLAS $\pi N$ photoproduction
data~\cite{CLAS:2009tyz} are shown by the blue triangles and squares, respectively.}
\label{n1520el_npi_pipimp}
\end{center}
\end{figure*}

In order to compare results for the $N(1440)1/2^+$ and $N(1520)3/2^-$ electrocouplings in the $\pi^+\pi^-p$ electroproduction 
channel with the values from the analysis of $\pi N$ electroproduction, we must use common decay branching fractions to these final
states for each resonance. Within the JM23 model, the sum of the branching fractions into $\pi N$ and $\pi \pi N$ accounts for almost
100\% of the total decay widths of the $N(1440)1/2^+$ and $N(1520)3/2^-$. Since the $\pi N$ exclusive electroproduction channels are 
the most sensitive to contributions from the $N(1440)1/2^+$ and $N(1520)3/2^-$, we re-evaluated the branching fraction for the decay 
to the $\pi \pi N$ final states $BF(\pi \pi N)_{corr}$ as

\begin{equation}
\label{bnpipi}
BF(\pi \pi N)_{corr}=1-BF(\pi N).
\end{equation} 
For these resonance decays to $\pi \pi N$, it turns out that the estimated branching fractions $BF(\pi \pi N)_{corr}$ 
are slightly ($<$10\%) different with respect to those obtained from the $\pi^+\pi^-p$ fit ($BF(\pi \pi N)_0$). Therefore, we 
multiplied the $\pi \Delta$ and $\rho p$ hadronic decay widths of the $N(1440)1/2^+$ and $N(1520)3/2^-$ from the $\pi^+\pi^-p$ fit by 
the ratio $\frac{BF(\pi \pi N)_{corr}}{BF(\pi \pi N)_0}$. The electrocouplings obtained were then multiplied by the correction factors 
\begin{equation}
\label{bnpipi1}
C_{hd}=\sqrt{\frac{BF(\pi \pi N)_{corr}}{BF(\pi \pi N)_0}}
\end{equation} 
in order to keep the resonant parts and the computed $\pi^+\pi^-p$ differential cross sections unchanged under the re-scaling of the
resonance hadronic decay parameters described above.

A special procedure was developed for the evaluation of the transverse $A_i$ ($i = 1/2$, $3/2$) and longitudinal $S_i$ ($i = 1/2$)
electrocouplings analyzing the results from independent fits of the electroproduction cross sections in the three 
$W$ intervals with electrocouplings and uncertainties $A_{i,j} \pm \delta A_{i,j}$ and $S_{i,j} \pm \delta S_{i,j}$, where
the index $j = 1 \to 3$ is the $W$ interval. First, we found the overlap range for the electrocouplings [$A_i^{min}-A_i^{max}$] 
($i = 1/2$, $3/2$) and [$S_i^{min}-S_i^{max}$] ($i = 1/2$) from the data fit in the three $W$ intervals

\begin{eqnarray}
\label{el_ranges}
A_i^{min} = min[A_{i,j}+\delta A_{i,j}] \\ \nonumber
S_i^{min} = min[S_{i,j}+\delta S_{i,j}]  \\ \nonumber
A_i^{max} = max[A_{i,j}-\delta A_{i,j}] \\ \nonumber
S_i^{max} = max[S_{i,j}-\delta S_{i,j}].  \nonumber
\end{eqnarray}
Within these ranges the best data description was achieved in all $W$ intervals. Consequently, the mean values for $A_i$ and $S_i$ 
were redefined as:
\begin{eqnarray}
\label{el_mean}
A_i=\frac{A_i^{min}+A_i^{max}}{2}, ~ (i=1/2,3/2) \\ \nonumber
S_i=\frac{S_i^{min}+S_i^{max}}{2}, ~ (i=1/2). \\ \nonumber
\end{eqnarray}

There are three sources of uncertainties in the evaluation of $A_i$ and $S_i$ in each of the three $W$ intervals: a) the range of 
overlap between the electrocouplings determined from the data fit defined in Eq.(\ref{el_ranges}), b) the root mean square (RMS) for 
the mean values of the determined electrocouplings, {\it i.e.} RMS [$A_{i,j}$] and RMS [$S_{i,j}$], and c) the differences between the 
redefined electrocouplings according to Eq.(\ref{el_mean}) and their average values obtained from the data fit. The total 
uncertainties were obtained as the quadrature sum of the contributions a) to c)
\begin{eqnarray}
\label{uncertainties}
\delta A_i=\sqrt{\frac{(A_i^{max}-A_i^{min})^2}{4}+(RMS[A_{i,j}])^2+\Delta A_i^2} \,  \\ \nonumber
\Delta A_i=\frac{A_i^{min}+A_i^{max}}{2}-\frac{\sum_{j=1,2,3} A_{i,j}}{3} \\ [0.5ex] \nonumber
\delta S_i=\sqrt{\frac{(S_i^{max}-S_i^{min})^2}{4}+(RMS[S_{i,j}])^2+\Delta S_i^2}\,  \\ 
\Delta S_i=\frac{S_i^{min}+S_i^{max}}{2}-\frac{\sum_{j=1,2,3} S_{i,j}}{3}.  \nonumber
\end{eqnarray}
The electrocouplings determined for the $N(1440)1/2^+$ and $N(1520)3/2^-$ are listed in Tables~\ref{p11el_final} and
\ref{d13el_final}, respectively. We consider these results as the final electrocouplings from the analysis of the $\pi^+\pi^-p$ 
electroproduction cross sections within the JM23 model.

\begin{table}
\begin{center}
\begin{tabular}{|c|c|c|}
\hline
$Q^2$ Interval,  & $A_{1/2} \times 1000$, & $S_{1/2} \times 1000$,    \\
 GeV$^2$          & GeV$^{-1/2}$        & GeV$^{-1/2}$        \\ \hline
 2.0-2.4          &  68.7 $\pm$ 6.6     & 7.6 $\pm$ 4.1  \\ \hline
 2.4-3.0          &  60.2 $\pm$ 4.9     & 11.8 $\pm$ 4.2  \\ \hline
 3.0-3.5          &  32.3 $\pm$ 7.5     & 8.7 $\pm$ 5.3  \\ \hline
 3.5-4.2          &  24.6 $\pm$ 3.6     & 11.6 $\pm$ 3.0  \\ \hline
 4.2-5.0          &  22.8 $\pm$ 4.3     & 3.3 $\pm$ 2.0  \\ \hline
\end{tabular}
\caption{$N(1440)1/2^+$ electrocouplings determined from the $\pi^+\pi^-p$ differential cross sections measured with the CLAS detector
\cite{CLAS:2017fja,Trivedi:2018rgo} in three $W$ intervals, 1.41--1.51~GeV, 1.46--1.56~GeV, and 1.51--1.61~GeV, for $Q^2$ from 
2.0--5.0~GeV$^2$ evaluated according to Eqs.(\ref{el_mean},\ref{uncertainties}).} 
\label{p11el_final}
\end{center}
\end{table}

\begin{table}
\begin{center}
\begin{tabular}{|c|c|c|c|}
\hline
$Q^2$ Interval,  & $A_{1/2} \times 1000$, & $S_{1/2} \times 1000$, & $A_{3/2} \times 1000$,  \\
 GeV$^2$          & GeV$^{-1/2}$        & GeV$^{-1/2}$  &  GeV$^{-1/2}$    \\ \hline
 2.0-2.4          &  -35.8 $\pm$ 4.8     & -3.8 $\pm$ 5.4 &  16.8 $\pm$ 2.2 \\ \hline
 2.4-3.0          &  -37.5 $\pm$ 5.2     & -4.1 $\pm$ 5.6 &  15.8 $\pm$ 3.2 \\ \hline
 3.0-3.5          &  -25.7 $\pm$ 3.5     & -0.1 $\pm$ 2.4 &  9.0 $\pm$ 3.5 \\ \hline
 3.5-4.2          &  -16.8 $\pm$ 5.2     & -3.3 $\pm$ 1.4 &  11.2 $\pm$ 4.3 \\ \hline
 4.2-5.0          &  -15.6 $\pm$ 6.7     & -0.5 $\pm$ 3.0 &  7.0 $\pm$ 4.1 \\ \hline
\end{tabular}
\caption{$N(1520)3/2^-$ electrocouplings determined from the $\pi^+\pi^-p$ differential cross sections measured with the CLAS detector
\cite{CLAS:2017fja,Trivedi:2018rgo} in three $W$ intervals, 1.41--1.51~GeV, 1.46--1.56~GeV, and 1.51--1.61~GeV, for $Q^2$ from 
2.0--5.0~GeV$^2$ evaluated according to Eqs.(\ref{el_mean},\ref{uncertainties}).} 
\label{d13el_final}
\end{center}
\end{table}

In Figs.~\ref{n1440el_npi_pipimp} (left) and \ref{n1520el_npi_pipimp} (left) we compare the electrocouplings of the $N(1440)1/2^+$ 
and $N(1520)3/2^-$ obtained from analyses of $\pi N$ electroproduction cross sections, beam, target, and beam-target asymmetries 
within the unitary isobar model and dispersion relation approach \cite{Aznauryan:2009mx,Aznauryan:2002gd} with the results available 
from analysis of the nine independent one-fold differential $\pi^+\pi^-p$ electroproduction cross sections within JM23. The analyses 
of the $\pi N$ and $\pi^+\pi^-p$ data were carried out with different $Q^2$-binning. For direct comparison, the electrocouplings 
obtained from $\pi N$ were interpolated over $Q^2$, so that the electrocouplings could be compared at the same $Q^2$-values. The 
comparison between all currently available electrocouplings for the $N(1440)1/2^+$ and $N(1520)3/2^-$ from CLAS $\pi N$ and 
$\pi^+\pi^-p$ electroproduction is shown in Figs.~\ref{n1440el_npi_pipimp} (right) and \ref{n1520el_npi_pipimp} (right). Overall, 
good agreement has been achieved between the electrocouplings of both the $N(1440)1/2^+$  and $N(1520)3/2^-$ determined from 
independent analyses of the $\pi N$ and $\pi^+\pi^-p$ data.

The $\pi N$ and $\pi^+\pi^-p$ electroproduction channels account for the largest part of the total meson electroproduction cross 
sections in the resonance region for $W < 2$~GeV. Their non-resonant amplitudes are different, however, the $N^*$ electrocouplings 
obtained from independent studies of these channels should be the same at each $Q^2$, since the resonance electroexcitation and 
hadronic decay amplitudes into the different final states should be independent. Hence, consistent results on the electrocouplings 
of the $N(1440)1/2^+$ and $N(1520)3/2^-$ deduced from $\pi N$ and $\pi^+\pi^-p$ observed within a broad range of $Q^2$ from the 
photon point to 5.0~GeV$^2$ validates the extraction of the $\gamma_vpN^*$ electrocouplings from the $\pi^+\pi^-p$ electroproduction 
data.

\begin{figure*}[htbp]
\begin{center}
\includegraphics[width=15.0cm]{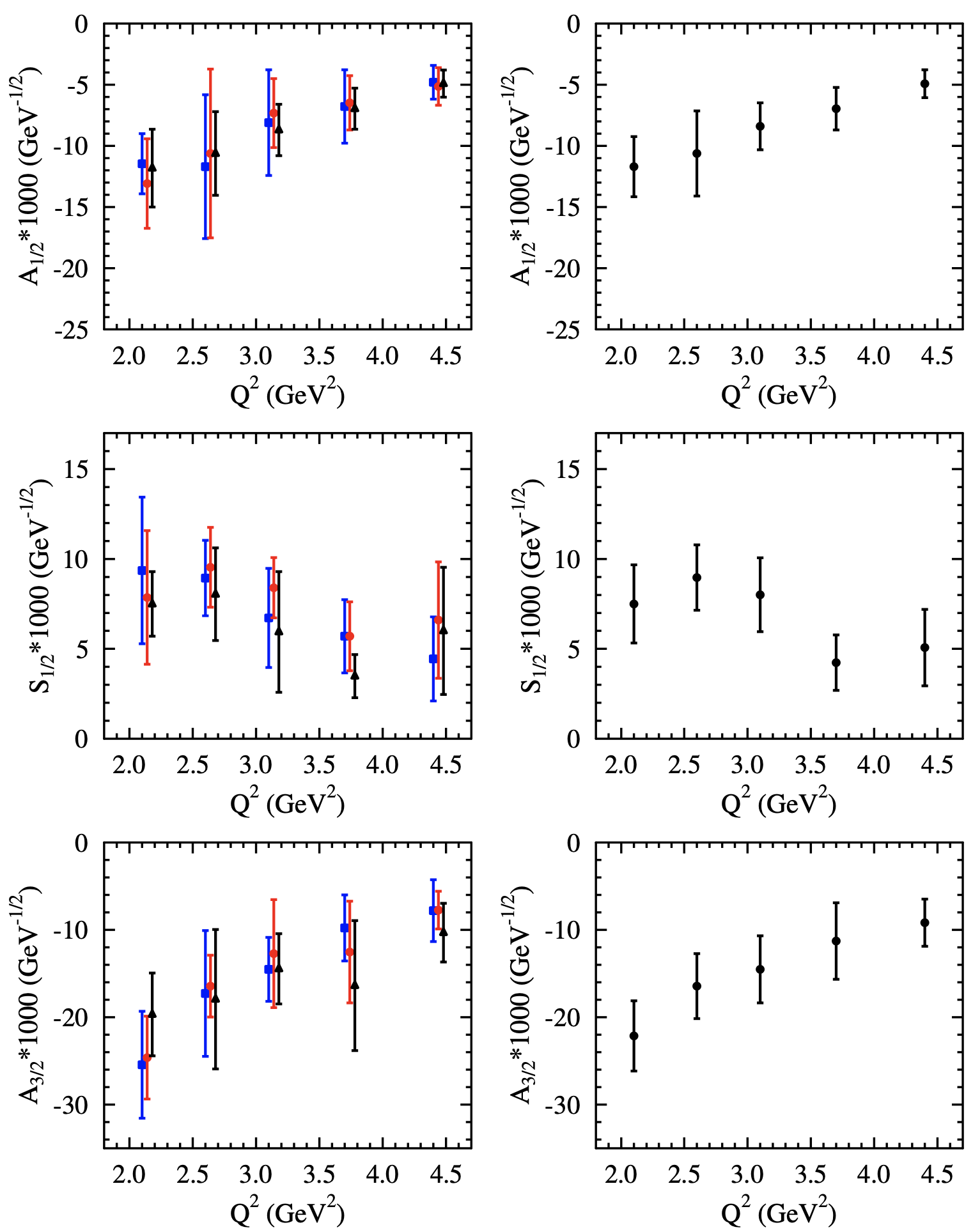}
\caption{(Left) $\Delta(1600)3/2^+$ electrocouplings deduced from independent fits of the $\pi^+\pi^-p$ differential cross sections 
carried out within three $W$ intervals, 1.46--1.56~GeV (blue squares), 1.51--1.61~GeV (red circles), and 1.56--1.66~GeV (black 
triangles), for $Q^2$ from 2.0--5.0~GeV$^2$. (Right) $\Delta(1600)3/2^+$ electrocouplings evaluated by combining the results from 
the three $W$ intervals as described in Section~\ref{n1440n1520}.}
\label{d1600el_3w_aver}
\end{center}
\end{figure*}

\subsection{Parameters for the $\Delta(1600)3/2^+$} 
\label{delta1600}

After the discovery of several new excited states of the proton from global multichannel analysis of exclusive meson photo- and
hadroproduction data~\cite{Burkert:2019kxy}, the status of the $\Delta(1600)3/2^+$ was elevated to a four-star firmly established
state~\cite{ParticleDataGroup:2022pth}. This resonance decays preferentially into $\pi\Delta$ and has been included in 
the JM23 model. 

\begin{table}
\begin{center}
\begin{tabular}{|c|c|c|c|c|c|} \hline
$W$ Inter-    & $Q^2$ Inter- & Mass,          &$\Gamma_{tot}$, &$\Gamma_{\pi\Delta}$, &  $BF_{\pi\Delta}$, \\
val, GeV      & val, GeV$^2$ & GeV            &  GeV           &  GeV                 & \%                  \\ \hline
1.46-1.56     & 2.0-3.5      & 1.55$\pm$0.014 & 244$\pm$21    &  154$\pm$21 & 50-78 \\ \hline
1.51-1.61     & 2.0-3.5      & 1.57$\pm$0.018 & 259$\pm$21    &  169$\pm$22 & 52-81 \\ \hline
1.56-1.66     & 2.0-3.5      & 1.57$\pm$0.042 & 256$\pm$33    &  166$\pm$34 & 46-90 \\ \hline
1.46-1.56     & 3.0-5.0      & 1.56$\pm$0.030 & 249$\pm$37    &  158$\pm$37 & 42-92 \\ \hline
1.51-1.61     & 3.0-5.0      & 1.56$\pm$0.030 & 249$\pm$34    &  158$\pm$34 & 44-89 \\ \hline
1.56-1.66     & 3.0-5.0      & 1.58$\pm$0.039 & 263$\pm$29    &  172$\pm$29 & 49-86 \\ \hline
  PDG          & PDG         & 1.50-1.64      & 200-300       &  172$\pm$29 & 73-83 \\ \hline
\end{tabular}
\caption{Mass and total/partial decay widths of the $\Delta(1600)3/2^+$ into $\pi\Delta$ determined from the fit of  $\pi^+\pi^-p$
electroproduction cross sections carried out independently within two intervals in $Q^2$ and within three overlapping intervals in 
$W$. The PDG parameters are listed in the bottom row.}
\label{hadr_d1600} 
\end{center}
\end{table}

The electrocouplings, mass, and total/partial hadronic decay widths of the $\Delta(1600)3/2^+$ have been determined for the first
time from independent analysis of the $\pi^+\pi^-p$ differential cross sections within three overlapping $W$ intervals, 1.46--1.56~GeV, 
1.51--1.61~GeV, and 1.56--1.66~GeV, for $Q^2$ from 2.0--3.5~GeV$^2$ and from 3.0--5.0~GeV$^2$. The $\Delta(1600)3/2^+$ contributes
substantially to each of these intervals.

The mass and hadronic decay parameters of the $\Delta(1600)3/2^+$ determined from independent fits within the six overlapping 
($W$,$Q^2$) bins listed in Table~\ref{hadr_d1600} are consistent within their uncertainties. They are also in good agreement with
reported PDG values~\cite{ParticleDataGroup:2022pth}. The masses and hadronic decay widths of $N^*$s excited in the $s$-channel
for $\gamma_v p$ interactions should be $Q^2$-independent, since the resonance electroexcitation and hadronic decay amplitudes are
independent. The resonance hadronic decay widths obtained from the data fit within the three $W$ intervals should also be the same 
since the corresponding hadronic decay amplitudes are defined at the resonant point $W=M_{N^*}$. Therefore, the results in 
Table~\ref{hadr_d1600} provide evidence for the manifestation of the $\Delta(1600)3/2^+$ as an $s$-channel resonance seen 
in $\pi^+\pi^-p$ electroproduction. Furthermore, the pronounced $Q^2$-evolution observed in the non-resonant contributions makes
interpretation of the $\Delta(1600)3/2^+$ as a singularity of the non-resonant amplitudes or a dynamically generated resonance 
unlikely. The successful description of the $\pi^+\pi^-p$ electroproduction data within the ($W$,$Q^2$) bins listed in 
Table~\ref{hadr_d1600} achieved with $W$- and $Q^2$-independent $\Delta(1600)3/2^+$ mass and total/partial hadronic decay widths has 
also demonstrated the capability of the JM23 model for the evaluation of the resonant contributions from this state.  

\begin{table}
\begin{center}
\begin{tabular}{|c|c|c|c|}
\hline
$Q^2$ Interval,  & $A_{1/2} \times 1000$, & $S_{1/2} \times 1000$, & $A_{3/2} \times 1000$,  \\
 GeV$^2$          & GeV$^{-1/2}$        & GeV$^{-1/2}$  &  GeV$^{-1/2}$    \\ \hline
 2.0-2.4          &  -11.7 $\pm$ 2.5     & 7.5 $\pm$ 2.2 &  -22.1 $\pm$ 4.0 \\ \hline
 2.4-3.0          &  -10.6 $\pm$ 3.5     & 9.0 $\pm$ 1.8 &  -16.4 $\pm$ 3.7 \\ \hline
 3.0-3.5          &  -8.4 $\pm$ 1.0     &  8.0 $\pm$ 2.1 &  -14.5 $\pm$ 3.8 \\ \hline
 3.5-4.2          &  -7.0 $\pm$ 1.7     & 4.2 $\pm$ 1.5 &  -11.3 $\pm$ 4.4 \\ \hline
 4.2-5.0          &  -4.9 $\pm$ 1.1     & 5.1 $\pm$ 2.1 &  -9.2 $\pm$ 2.7 \\ \hline
\end{tabular}
\caption{$\Delta(1600)3/2^+$ electrocouplings determined from the $\pi^+\pi^-p$ differential cross sections measured with the CLAS 
detector in three $W$ intervals, 1.46--1.56~GeV, 1.51--1.61~GeV, and 1.56--1.66~GeV, for $Q^2$ from 2.0--5.0~GeV$^2$ evaluated as 
described in Section~\ref{n1440n1520}.} 
\label{p33el_final}
\end{center}
\end{table}

The procedure for the extraction of the $\Delta(1600)3/2^+$ electrocouplings is similar to that used for the $N(1440)1/2^+$ and
$N(1520)3/2^-$ described in Section~\ref{fit_strategy}. As starting values for the $\Delta(1600)3/2^+$ electrocouplings, we 
explored a $\pm$50\% range around the values predicted by CSM~\cite{Lu:2019bjs}. Under this variation, the $\pi^+\pi^-p$
differential cross sections computed within JM23 are spread within a range that overlaps the measured differential cross 
sections for the dominant part of the CLAS data points~\cite{CLAS:2017fja,Trivedi:2018rgo}. The extracted $\Delta(1600)3/2^+$ 
electrocouplings within each of the three $W$ intervals are shown in Fig.~\ref{d1600el_3w_aver} (left). The non-resonant amplitudes 
in the three $W$ intervals are different, however, the determined $\Delta(1600)3/2^+$ electrocouplings are the same within their 
uncertainties. This success solidifies the evidence for the extraction of the $\Delta(1600)3/2^+$ electrocouplings. The final results 
for the $\Delta(1600)3/2^+$ electrocouplings were determined by combining the results obtained from the data fits in the three $W$ 
intervals using the procedure described in Section~\ref{n1440n1520}. They are listed in Table~\ref{p33el_final} and shown in 
Fig.~\ref{d1600el_3w_aver} (right). Currently, the CLAS $\pi^+\pi^-p$ electroproduction cross sections are the only data from
which electrocouplings of this state have become available. Extraction of the $\Delta(1600)3/2^+$ electrocouplings from $\pi^0p$
electroproduction data~\cite{CLAS:2019cpp, CLAS:2021cvy} will be the next important step in the exploration of the structure of this
state.

\begin{figure*}[htbp]
\begin{center}
\includegraphics[width=12.5cm]{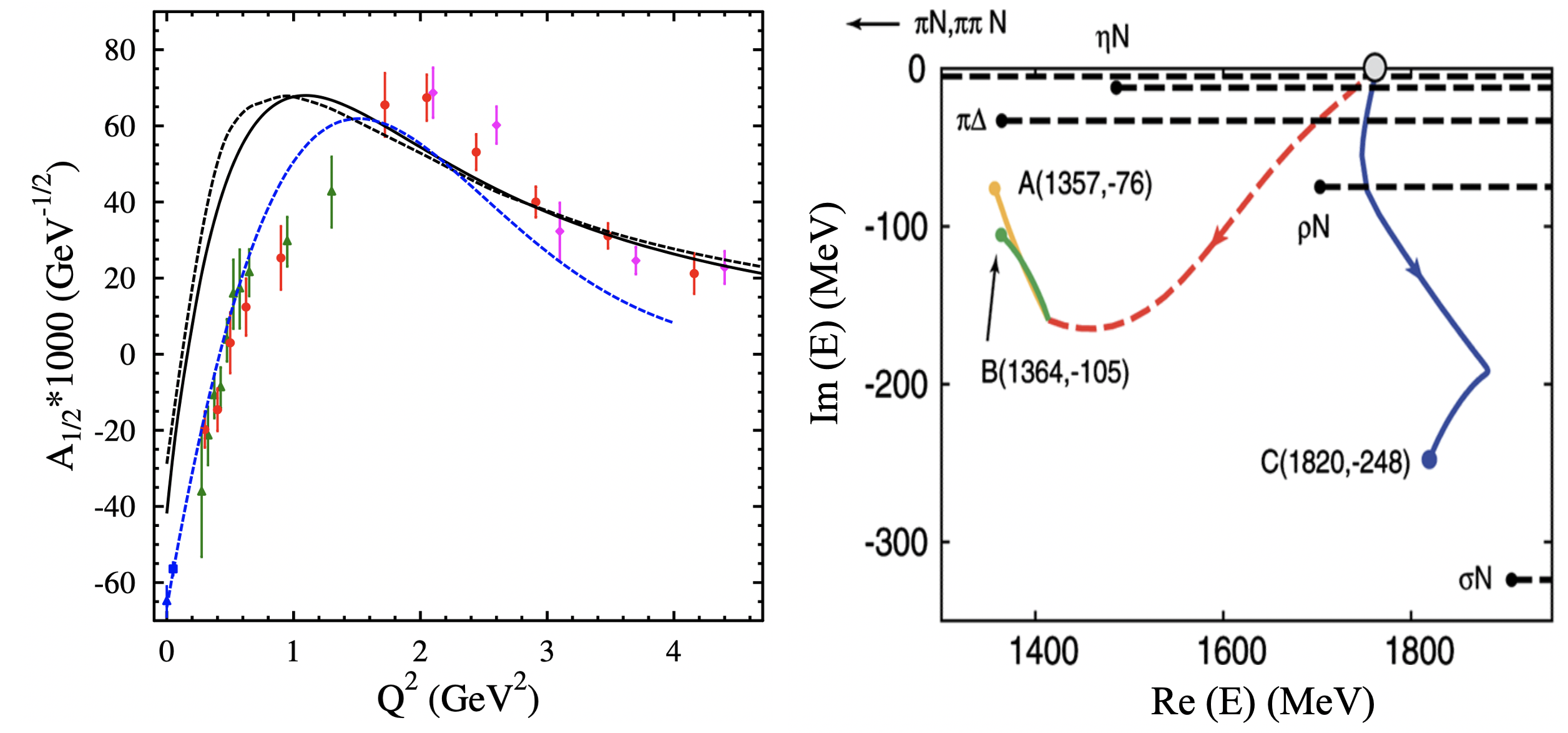}
\caption{(Left) $N(1440)1/2^+$ $A_{1/2}$ electrocouplings determined from studies of $\pi N$ electroproduction (red circles)
\cite{Aznauryan:2009mx} and from $\pi^+\pi^-p$ electroproduction for $Q^2 < 1.5$~GeV$^2$ (green triangles)~\cite{Mokeev:2012vsa,
Mokeev:2015lda} and for $Q^2$ from 2.0--5.0~GeV$^2$ available from this work (magenta diamonds). The electrocoupling description 
within CSM~\cite{Segovia:2015hra}, employing a momentum dependent dressed quark mass deduced from the QCD Lagrangian, is shown by 
the black solid line. The descriptions achieved within light-front quark models are shown that a) implement a phenomenological 
momentum-dependent dressed quark mass~\cite{Aznauryan:2018okk} (black dashed line) and b) account for both an inner core of three 
constituent quarks and an external meson-baryon cloud \cite{Obukhovsky:2011sc} (blue dashed line). (Right) The evolution of the
$N(1440)1/2^+$ complex pole mass available from the analysis of meson photo- and hadroproduction data within the Argonne-Osaka 
coupled-channel approach~\cite{Suzuki:2009nj,Aznauryan:2012ba} for running values of the meson-baryon couplings from zero (corresponding 
to the bare quark-core mass on the real energy axis shown by the shaded gray circle) to the finite values determined from the data. The
mass of the observed $N(1440)1/2^+$ is determined by the two poles in the complex energy plane labeled on the graph as A and B. The 
colored lines show the pole movement and splitting as the meson-baryon couplings increase. The horizontal dashed lines show the cuts 
owing to the opening of the quasi-two-body channels with unstable hadrons.} 
\label{p11_lf_csm}
\end{center}
\end{figure*}

\begin{figure}[htbp]
\begin{center}
\includegraphics[width=8.0cm]{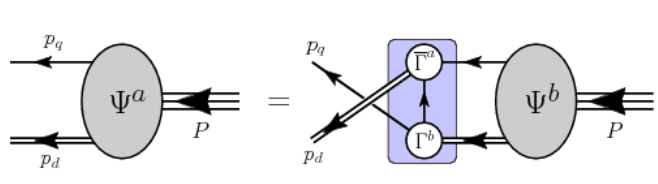}
\vspace{-1mm}
\caption{Faddeev equation for computation of the masses and wave functions of the quark core of the ground and excited states of the 
nucleon. The kernel for the matrix-valued integral equations is represented by the blue area.} 
\label{Faddeev_eq}
\end{center}
\end{figure}

\section{Insight into Nucleon Resonance Structure}
\label{impact_theor}

The new results on the $\gamma_vpN^*$ electrocouplings available from the analysis of the CLAS $\pi^+\pi^-p$ electroproduction data 
for $Q^2$ from 2.0--5.0~GeV$^2$, together with the previously available results from $\pi N$ and $\pi^+\pi^-p$ electroproduction off
protons for $Q^2 < 1.5$~GeV$^2$, provide important input needed to check theory predictions on the structure of $N^*$ states and 
their emergence from QCD~\cite{Brodsky:2020vco,Barabanov:2020jvn,Aznauryan:2012ba, Aznauryan:2011qj}. In this Section, we describe new
opportunities for the exploration of the structure of the $N(1440)1/2^+$, $N(1520)3/2^-$, and $\Delta(1600)3/2^+$ provided by the results
on their electrocouplings. 

\subsection{$N(1440)1/2^+$ Resonance Structure}
\label{roper}

The CLAS results on the electrocouplings of the $N(1440)1/2^+$ have been described for $Q^2$ from 2.0--5.0~GeV$^2$ within 
different approaches. The relativistic light-front quark model~\cite{Aznauryan:2018okk} and the CSM~\cite{Segovia:2015hra} are shown
as representative examples in Fig.~\ref{p11_lf_csm} (left) for the description of its $A_{1/2}$ electrocoupling.

The CSM approach provides a good description in the range of $Q^2$ from 2.0--5.0~GeV$^2$ of the electrocouplings of the $N(1440)1/2^+$ 
as a bound quark$+$diquark system in its first radial excitation~\cite{Segovia:2015hra}. Within this approach, the momentum dependence 
of the dressed quark and gluon masses has been evaluated from the solution of the QCD equations of motion for the quark and gluon fields.
The gluon self-interaction encoded in the QCD Lagrangian underpins the emergence of the dynamically generated gluon mass, which at 
distances on the order of the hadron size, approaches the mass scale of $\approx$0.4~GeV. This process is responsible for the sharp 
increase of the QCD running coupling $\alpha_s/\pi$ at distances where perturbative QCD evolves into the strongly coupled QCD (sQCD)
regime~\cite{Roberts:2021xnz,Roberts:2020hiw}. At quark momenta below 2~GeV, where $\alpha_s/\pi$ increases rapidly and becomes 
comparable with unity, the energy stored in the gluon field is transformed into the momentum dependence of the dynamically generated
dressed quark mass, which increases rapidly with increasing distance (or inverse quark momentum) and approaches the mass scale of
$\approx$0.4~GeV at quark momenta $<$ 0.5~GeV. The dressed quarks with momentum-dependent masses deduced from QCD are treated as the
building blocks for the quark core of the ground and excited state nucleons. Their masses and wave functions are obtained from the 
solution of the Faddeev equations for three dressed quarks in the approximation of a quark$+$diquark kernel 
\cite{Segovia:2014aza,Segovia:2015hra,Barabanov:2020jvn}. 

The diquark correlations employed in CSM are different in comparison with the rigid diquarks of constituent quark models. In the
CSM approach, diquarks represent correlated quark pairs, whose correlation amplitudes are computed as the solution of the 
Bethe-Salpeter equation with the kernel for the two-dressed-quark interaction mediated by dressed gluon exchange starting from the
QCD Lagrangian. The CSM diquark is a dynamical object that interacts with the corresponding third quark, forming a new correlated 
diquark pair, as shown in Fig.~\ref{Faddeev_eq}. The masses of the ground and excited states of the nucleon of a given spin-parity 
$J^P$ have been obtained as poles in the respective $J^P$ partial waves of the Faddeev amplitude of the three dressed quarks from the
solution of the Faddeev equations depicted in Fig.~\ref{Faddeev_eq}. The wave functions for the ground and excited states of the 
nucleon were obtained from the Faddeev amplitude residues at the pole positions. The electrocouplings are evaluated considering the 
virtual photon interaction with the electromagnetic currents of the dressed quark and diquark system for the transitions between 
diquarks of the same or different spin-parities, and account for the virtual photon interaction at the vertex describing diquark
decay/recombination to/from the pair of uncorrelated quarks shown by the blue shadowed area in Fig.~\ref{Faddeev_eq}
\cite{Segovia:2014aza}. 

Within the light-front quark model~\cite{Aznauryan:2012ec,Aznauryan:2018okk}, the $N(1440)1/2^+$ is treated as a bound system of three
constituent quarks in their first radial excitation. The momentum dependence of the constituent quark mass has been employed in order 
to reproduce the experimental results on the nucleon elastic form factors. With the same momentum dependence of the constituent quark 
mass the model has succeeded in providing a reasonable description of the electrocouplings of all $N^*$ states in the mass range up to
1.6~GeV.

As shown in Fig.~\ref{p11_lf_csm} (left), both CSM and the light-front quark model describe the $N(1440)1/2^+$ electrocouplings for 
$Q^2$ from 2.0--5.0~GeV$^2$ by employing a momentum-dependent quark mass with virtually coincident predictions. This success 
demonstrates strong support for quarks with running mass as active structural components in the ground and excited states of the 
nucleon at distances where the contributions from the quark core become the largest, which occurs at $Q^2 \gtrsim 2$~GeV$^2$ for the 
$N(1440)1/2^+$. However, both the CSM~\cite{Segovia:2015hra} and the light-front quark model of Ref.~\cite{Aznauryan:2018okk} fail to 
reproduce the $N(1440)1/2^+$ electrocouplings for $Q^2 < 1$~GeV$^2$. This failure points to additional contributions to the structure
relevant at distances on the order of the baryon size. These contributions arise from the meson-baryon cloud. The light-front quark 
model of Ref.~\cite{Obukhovsky:2011sc}, which takes into account the contributions from both the meson-baryon cloud and the quark core 
to the structure of the $N(1440)1/2^+$, provides a much better description of the data at low $Q^2$, while retaining a reasonable 
description at higher $Q^2$. This feature explains the success of the models that account for only the meson-baryon degrees of freedom 
in describing the electrocouplings of the $N(1440)1/2^+$ for $Q^2 < 1$~GeV$^2$~\cite{Bauer:2014cqa,Krehl:1999km,Speth:2000zf}. 

Taking into account the contributions from the quark core and meson-baryon cloud allowed for the resolution of the long-standing 
puzzle on the ordering of the masses of the radial and orbital excitations in the $N^*$ spectrum. Most quark models with quark 
interactions mediated by gluon exchange predict the mass of the first radial excitation of three quarks above the mass of the first 
orbital excitation. The experimental results, in contrast with these quark model expectations, revealed that the mass of the 
$N(1440)1/2^+$ (1.440~GeV), a state that represents the first radial excitation of three quarks, is below that of the states belonging to 
the [70,1$^-$] SU(6) spin-flavor super-multiplet (1.520~GeV and 1.535~GeV for the lightest states) and thus should be expected as the 
first orbital excitation of three quarks. Coupled-channel analyses of the meson photo- and hadroproduction data carried out by the
Argonne-Osaka group~\cite{Suzuki:2009nj,Aznauryan:2012ba} (see Fig.~\ref{p11_lf_csm} (right)) revealed that in the limit of zero
meson-baryon coupling corresponding to the contribution from just the bare quark core, the mass of 1.76~GeV of the $N(1440)1/2^+$ 
radial excitation would be above the masses of the lightest $N^*$ in the [70,1$^-$] super-multiplet, and hence in agreement with the
quark model expectations. As the meson-baryon couplings increase toward the values established in the data analysis, the single pole 
on the real energy axis moves into the complex energy plane and eventually splits into two poles A and B related to the $N(1440)1/2^+$
properties, while the third pole C moves toward the mass range above 1.7~GeV as shown in Fig.~\ref{p11_lf_csm} (right). Therefore, the
meson-baryon dressing is responsible for the shift of the bare quark core mass down to the value of the measured mass, below the lightest
states in the [70,1$^-$] super-multiplet. 

The CLAS results on the $N(1440)1/2^+$ electrocouplings available over a broad range of $Q^2 < 5.0$~GeV$^2$ (see Fig.~\ref{p11_lf_csm})
have revealed its structure as an interplay between the inner core of three dressed quarks in their first radial excitation 
augmented by the external meson-baryon cloud. The meson-baryon degrees of freedom are most relevant for $Q^2 \lesssim 1$~GeV$^2$, while 
photons of virtualities $Q^2 \gtrsim 2$~GeV$^2$ interact mostly with the quark core. Therefore, higher $Q^2$ studies are preferential 
for the exploration of the quark degrees of freedom in the structure of this state. Thus studies of the $N^*$ electrocouplings over a 
broad $Q^2$ range are necessary to establish the relevant degrees of freedom in their structure and to shed light on their 
distance-dependent evolution.

The CSM analysis has provided predictions for the quark core wave function of the $N(1440)1/2^+$ in terms of quark$+$diquark
configurations of certain values of the orbital angular momentum of the quark relative to the diquark, as well as for certain 
isospin and spin-parity values for the diquarks~\cite{Chen:2017pse}. The Faddeev equations are Poincar{\'e} covariant and 
their solution is available in any reference frame. For comparison with the quark model expectations, the quark$+$diquark content 
in the analysis of Ref.~\cite{Chen:2017pse} was evaluated in the resonance rest frame.

\begin{figure}[htbp]
\begin{center}
\includegraphics[width=7.1cm]{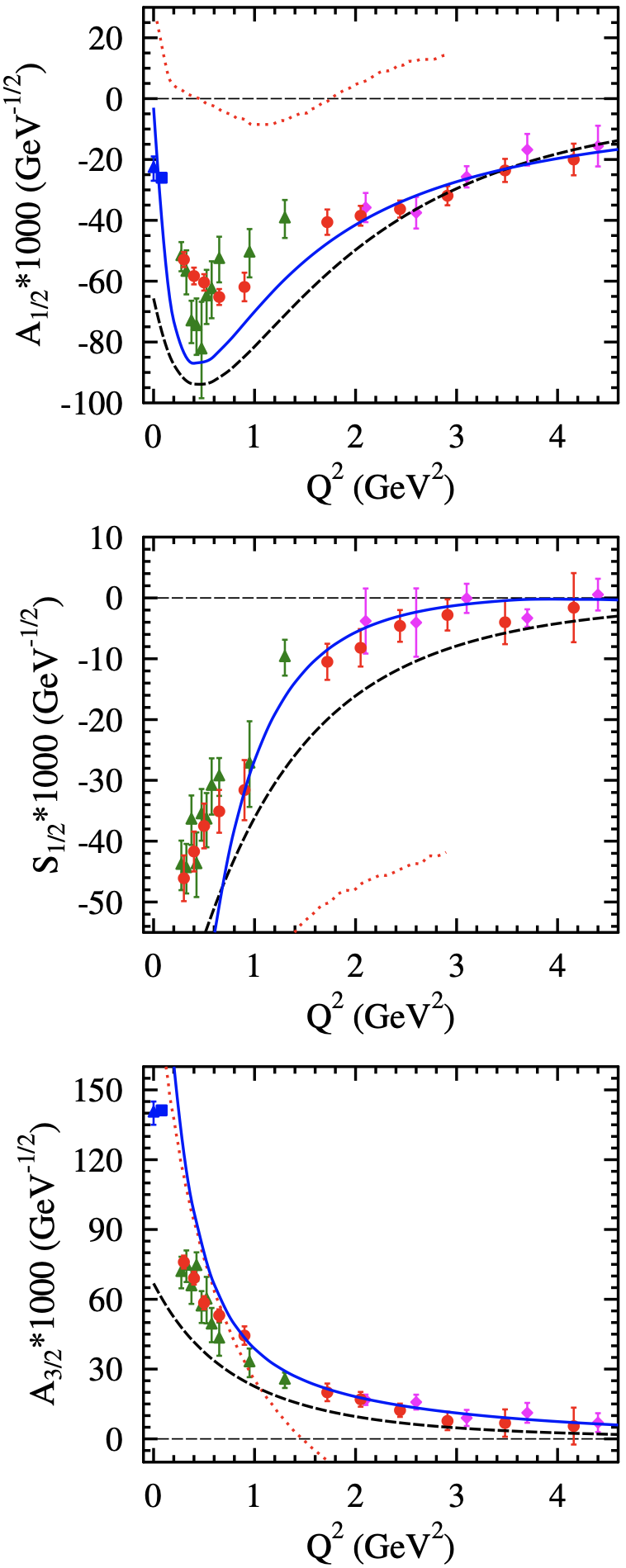}
\vspace{-2.mm}
\caption{$N(1520)3/2^-$ electrocouplings determined from studies of $\pi N$ electroproduction (red circles)~\cite{Aznauryan:2009mx},
and from $\pi^+\pi^-p$ electroproduction for $Q^2 < 1.5$~GeV$^2$(green triangles)~\cite{Mokeev:2012vsa, Mokeev:2015lda} and for $Q^2$
from 2.0--5.0~GeV$^2$ available from this work (magenta diamonds): $A_{1/2}(Q^2)$ (top), $S_{1/2}(Q^2)$ (middle), and $A_{3/2}(Q^2)$ 
(bottom). The electrocoupling descriptions achieved within the light-front quark model that includes only the three-quark configuration
expected for the SU(6) assignment of the $N(1520)3/2^-$~\cite{Capstick:1994ne} are shown by the dotted red lines. The results of the 
light-front quark model~\cite{Aznauryan:2018okk} that employs three quark configuration mixing and a phenomenological parameterization 
for the running quark mass and from the hypercentral constituent quark model~\cite{Giannini:2015zia} are shown by the solid blue and 
dashed black curves, respectively.} 
\label{d13_theor_exp}
\end{center}
\end{figure}

Evaluations of the contributions from quark$+$diquark configurations into the resonance mass within CSM showed that the dominant part 
of the bare quark core mass of the $N(1440)1/2^+$ is created by a configuration with a scalar diquark of spin-parity $J^P = 0^+$ and
relative orbital angular momentum $L = 0$ of the quark. This fully relativistic finding based on the QCD Lagrangian is in good 
agreement with expectations from the majority of constituent quark models~\cite{Giannini:2015zia,Capstick:2000qj,Obukhovsky:2019xrs,
Aznauryan:2012ba,Aznauryan:2012ec}. However, the evaluation of the contributions from the quark$+$diquark configurations in the 
$N(1440)1/2^+$ wave function has revealed a more complex pattern: a) the contributions from scalar $J^P = 0^+$ and axial-vector 
diquarks become comparable and b) higher orbital angular momenta of the quark also contribute to the wave function. Therefore, studies 
of only the spectrum of $N^*$ states have limited sensitivity to their structure. Any statement on $N^*$ structure based solely on
analysis of the $N^*$ spectrum is tenuous as such studies only account for the wave function behavior at distances comparable 
with the hadron size. The full complexity of the $N^*$ wave function can be mapped out from the results on the $Q^2$-evolution of the
electrocouplings as they are sensitive to all configurations contributing to the wave function and their distance-dependent evolution.

Lowest-order Chebyshev projections for all configurations contributing to the $N(1440)1/2^+$ quark core wave function evaluated within
CSM~\cite{Chen:2017pse} revealed zeros in their dependencies on the relative quark$+$diquark momentum. This serves as evidence for a
radial excitation in the three quark system. Again, this finding, obtained under direct connection to QCD, explains the success of the
quark model results \cite{Giannini:2015zia,Capstick:2000qj,Obukhovsky:2019xrs,Aznauryan:2012ba,Aznauryan:2012ec} for the 
quark core of the $N(1440)1/2^+$ as a radial excitation of the three quark system. The good description of the $N(1440)1/2^+$ 
electrocouplings obtained from the $\pi^+\pi^-p$ electroproduction data of this work and from $\pi N$ electroproduction for $Q^2$ from 
2.0--5.0~GeV$^2$ achieved with CSM supports these predictions on its quark core wave function.

\subsection{$N(1520)3/2^-$ Resonance Structure}
\label{d13_state}

The detailed and consistent information on the electrocouplings of the $N(1520)3/2^-$ available for $Q^2 < 5.0$~GeV$^2$ from the 
studies of $\pi N$ and $\pi^+\pi^-p$ electroproduction sheds light on the relevant degrees of freedom in its structure. From its 
SU(6) assignment, the quark core of the $N(1520)3/2^-$ can be described as three quarks in the octet flavor SU(3)-multiplet of mixed 
permutation symmetry with spin $S = 1/2$ and orbital angular momentum $L = 1$~\cite{Giannini:2015zia, Capstick:1994ne}. The predicted 
$N(1520)3/2^-$ electrocouplings accounting for only this three-quark configuration have become available from the relativistic quark 
model~\cite{Capstick:1994ne} and are shown in Fig.~\ref{d13_theor_exp} by the dotted red lines. These predictions are in strong 
disagreement with the experimental results. These significant discrepancies suggest that the structure of the $N(1520)3/2^-$ quark 
core is more complex than expected from its SU(6) assignment.

\begin{figure}[htbp]
\begin{center} 
\includegraphics[width=7.5cm]{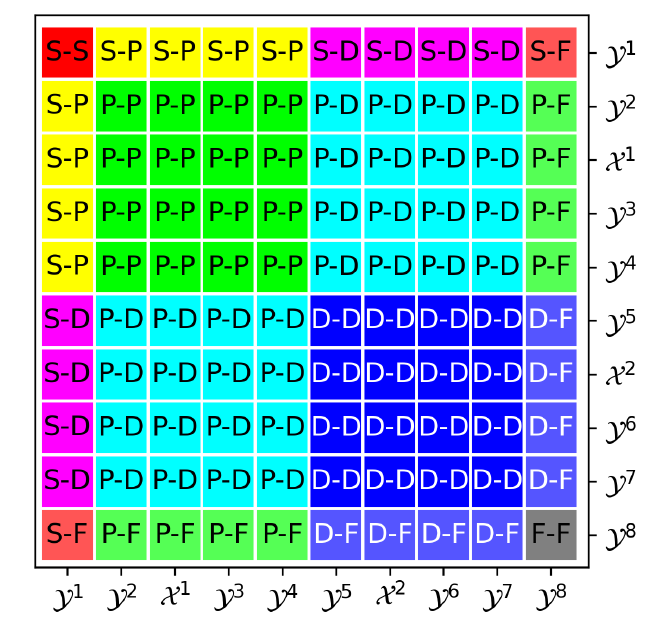}
\vspace{-1.5mm}
\includegraphics[width=7.5cm]{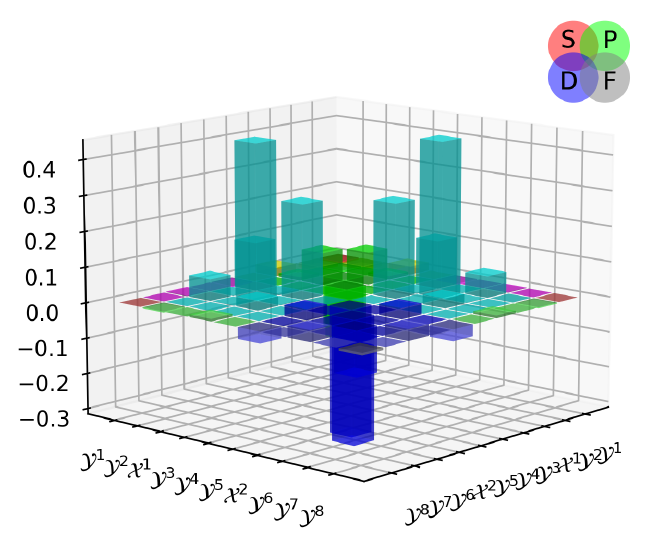}
\vspace{-1mm}
\caption{(Top) Color map for the contributions to the $N(1520)3/2^-$ wave function from quark$+$diquark configurations with $L$ and 
$L'$ orbital angular momenta in the canonical normalization constant of the $N^*$ wave functions accounting for $S$, $P$, $D$, and 
$F$ quark orbital angular momenta. Here, $L$ and $L'$ represent the quark orbital angular momenta in the resonance wave function and 
its conjugate, respectively. The axis labels represent the parts of the Faddeev amplitude that contain this information
\cite{Liu:2022nku}. (Bottom) The contributions from the pairs of quark$+$diquark configurations to the canonical normalization constant 
of the $N(1520)3/2^-$ quark core wave function computed within CSM~\cite{Liu:2022nku} in the resonance rest frame. The color code 
is shown in the top figure.} 
\label{d13 configurations}
\end{center}
\end{figure}

A reasonable description of the $N(1520)3/2^-$ electrocouplings has been achieved within the framework of the hypercentral constituent
quark model (hCQM)~\cite{Giannini:2015zia}. Within this approach, the space part of the $N(1520)3/2^-$ quark core wave function in the
resonance rest frame is described by six coordinates consisting of two solid angles $\Omega_{\rho}$, $\Omega_{\lambda}$, hyperangle 
$\xi$, and hyperradius $x$. Relations between $\xi$, $x$, and the $\rho$ and $\lambda$ vectors that describe the three quark systems 
can be found in Ref.~\cite{Giannini:2015zia}. Both $\xi$ and $x$ are determined by the coordinates of the three quarks. The
wave function of the three-quark system expressed in terms of the hypercentral coordinates effectively accounts for the interactions
between the three quarks, resulting in their binding into the quark core. The confining potential employed in the hCQM is a function of
$x$ and consists of two parts: a Coulomb term relevant for distances close to the pQCD regime $\propto 1/x$ and a linear confining part 
$\propto x$. The SU(6)-violating part of the $qq$-interaction consists of the hyperfine term stemming from vector particle exchange 
between the two quarks and the quark-flavor mixing term mediated by pseudoscalar meson exchange. Both terms are supported qualitatively 
by the CSM approach~\cite{Segovia:2014aza,Chen:2019fzn}. The dressed gluon exchange represents the vector particle exchange between 
two quarks, while pseudoscalar meson exchange could be traced back to dynamical chiral symmetry breaking at distances comparable with the 
baryon size scale. Both SU(6) violating terms and the linear confining term underlie the three-quark configuration mixing. The
electrocouplings of most $N^*$s have been computed within the hCQM for $Q^2 < 5.0$~GeV$^2$, keeping all model parameters at the values 
that fit the $N^*$ spectrum. The hCQM results on the $Q^2$-evolution of the $N(1520)3/2^-$ electrocouplings \cite{Giannini:2015zia} are 
shown in Fig.~\ref{d13_theor_exp} by the dashed black lines, and a reasonable description has been achieved for $Q^2 > 2.5$~GeV$^2$. 

The $N(1520)3/2^-$ electrocouplings have also been computed within the light-front quark model~\cite{Aznauryan:2018okk} and are shown
in Fig.~\ref{d13_theor_exp} by the solid blue lines. The major features of this model were highlighted in Section~\ref{roper}. The
electrocouplings of most $N^*$s in the mass range below 1.6~GeV have been evaluated within this model by employing the same 
momentum-dependent quark mass deduced from the fit of the electromagnetic nucleon elastic form factors. A reasonable description of 
the CLAS results for the $N(1520)3/2^-$ electrocouplings has been achieved for $Q^2$ from 1.5--5.0~GeV$^2$. The analysis of the CLAS
results on the $N(1440)1/2^+$ and $N(1520)3/2^-$ electrocouplings within the light-front quark model \cite{Aznauryan:2018okk} suggests 
that quarks with momentum-dependent mass represent the active degrees of freedom for the quark core structure of these states. Studies 
of the $N(1520)3/2^-$ electrocouplings within these models~\cite{Giannini:2015zia,Aznauryan:2018okk} have also demonstrated the important 
role of three-quark configuration mixing in the generation of this state. However, both of these models fail to describe the 
$N(1520)3/2^-$ electrocouplings for $Q^2 \lesssim 1$~GeV$^2$. These discrepancies are suggestive of substantial meson-baryon cloud 
contributions relevant at distances comparable with the baryon size. The CLAS results on the $N(1520)3/2^-$ electrocouplings show 
that its structure arises as a convolution between the external meson-baryon cloud and the inner core of three dressed quarks.

The contributions from different quark$+$diquark configurations to the quark core of the $N(1520)3/2^-$ have been evaluated within the
CSM approach~\cite{Liu:2022nku}. These studies demonstrated that 92\% of its quark core mass is generated by configurations with 
axial-vector diquarks of $J^P = 1^+$, while the combined contributions of the configurations with scalar $J^P = 0^+$ and axial-vector 
diquarks account for almost 100\% of the quark core mass, and with respect to the relative angular momentum the $P$-wave quark in the
quark$+$diquark rest frame generates 98\% of the $N(1520)3/2^-$ quark core mass. These CSM evaluations demonstrate that the dominant 
part of the quark core mass of this state is created by quark$+$diquark configurations consistent with its SU(6) assignment as the 
first orbital excitation of three quarks of $L = 1$ and total quark spin $S = 1/2$. 

However, analysis of the contributions from quark$+$diquark configurations to the Faddeev amplitude or wave function of the 
$N(1520)3/2^-$ reveals a more complex pattern that becomes evident in the studies of the contributions from the pairs of quark$+$diquark
configurations with orbital angular momenta $L$ and $L'$ in the canonical normalization constant for the resonance wave function
evaluated in its rest frame. Here, $L$ and $L'$ represent the quark orbital angular momenta in the resonance wave function and its
conjugate, respectively. The results are shown in Fig.~\ref{d13 configurations}. The major part of the $N(1520)3/2^-$ quark core wave
function is determined by interference between $P$- and $D$-waves and by the negative contribution of the $D$-wave. The contribution 
from the $P$-wave is smaller but non-negligible. These results qualitatively support the quark model findings on the substantial role 
of quark configuration mixing in the $N(1520)3/2^-$ quark core wave function. The CSM studies demonstrated that the spectroscopic
$N(1520)3/2^-$ mass is determined by just a $P$-wave quark$+$diquark configuration consistent with the SU(6) assignment for this state.
Therefore, studies of only the $N^*$ spectrum do not have sufficient sensitivity to elucidate the complexity of its structure. Indeed,
the resonance masses are determined by just the long wavelength part of the wave function. Instead, the electrocouplings are sensitive
to the contributions from different quark$+$diquark configurations, shedding light on the full complexity of their structure. 

Consequently, a comparison between the predicted CSM results on the $Q^2$-evolution of the $N(1520)3/2^-$ electrocouplings with the 
results determined deduced from the $\pi N$ and $\pi^+\pi^-p$ electroproduction data measured with CLAS provides a sensitive tool for 
validation of the CSM expectations on the complexity of quark$+$diquark configuration mixing in the $N(1520)3/2^-$ wave function 
obtained under a traceable connection to the QCD Lagrangian. The evaluations of the $N(1520)3/2^-$ electrocouplings within CSM are 
currently in progress.

\subsection{$\Delta(1600)3/2^+$ Resonance Structure}
\label{delta_1600}

\begin{figure}[h!]
\begin{center}
\vspace{-2mm}
\includegraphics[width=7.0cm]{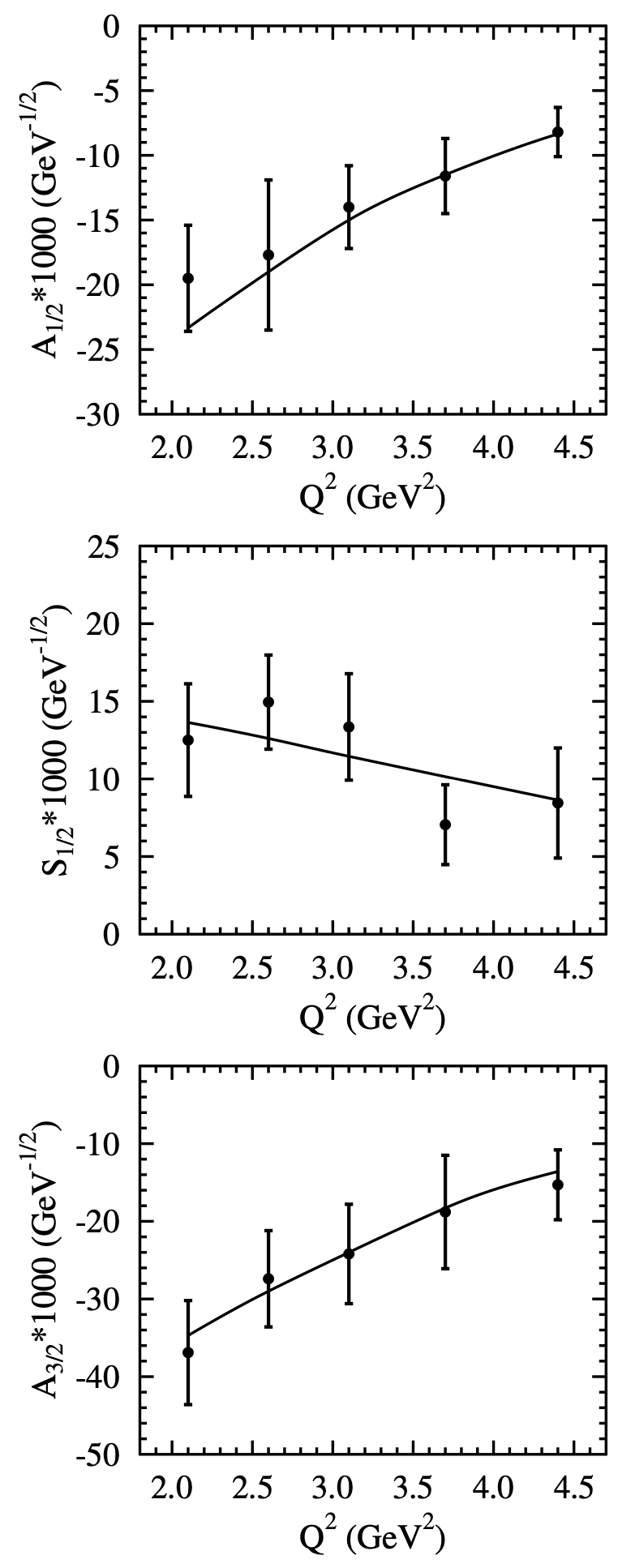}
\caption{$\Delta(1600)3/2^+$ electrocouplings obtained in this work: $A_{1/2}$ (top), $S_{1/2}$ (middle), and $A_{3/2}$ (bottom). For 
comparison with the CSM predictions~\cite{Lu:2019bjs} (solid black lines), the electrocouplings determined from the $\pi^+\pi^-p$ data
analysis have been divided by a factor of 0.6 to account for the missing contributions from the meson-baryon cloud in the CSM (see 
Section~\ref{delta_1600} for details).}
\label{p33_csm_exp}
\end{center}
\end{figure}

The predictions of the $Q^2$-evolution of the $\Delta(1600)3/2^+$ electrocouplings became available in 2019 within the CSM
\cite{Lu:2019bjs}. These evaluations employed the same momentum-dependent dressed quark mass deduced from the QCD Lagrangian and the 
same framework was used for solving the Faddeev equation to obtain the $\Delta(1600)3/2^+$ wave function as used previously in the
description of the $N(1440)1/2^+$ electrocouplings~\cite{Segovia:2015hra}. There are no additional free parameters in the computation 
of the $\Delta(1600)3/2^+$ electrocouplings. Furthermore, in 2019 no experimental results for the $\Delta(1600)3/2^+$ electrocouplings 
were available. These have been obtained for the first time from this work. These CSM predictions are shown in Fig.~\ref{p33_csm_exp} 
by the solid lines in comparison with the results determined from the data.

The studies of the electrocouplings have demonstrated that the $N^*$ wave functions are determined by the combined contributions from 
the quark core and meson-baryon cloud. The CSM evaluation of the $\Delta(1600)3/2^+$ electrocouplings accounts for the contribution from
only the quark core. Therefore, for comparison with the CSM expectations, the values deduced in this work should be divided by a factor 
to account for the contribution from only the three quark core component to the full resonance wave function normalization. This factor
should be the same for all three $\Delta(1600)3/2^+$ electrocouplings and $Q^2$-independent. We obtained this factor of 0.6 from the 
best description of the nine one-fold differential cross sections for $W$ from 1.46--1.66~GeV and $Q^2$ from 2.0--5.0~GeV$^2$ by varying
the common multiplicative factor for the three $\Delta(1600)3/2^+$ electrocouplings in the range from 0 to 1 with the other parameters of
the JM23 model that had been adjusted to fit the $\pi^+\pi^-p$ electroproduction cross section data. Figure~\ref{p33_csm_exp} shows the
electrocouplings from the right column of Fig.~\ref{d1600el_3w_aver} divided by a factor of 0.6 and compared with the CSM expectations. 
The results on the $Q^2$-evolution of the $\Delta(1600)3/2^+$ electrocouplings deduced from the $\pi^+\pi^-p$ electroproduction data in
this work have confirmed the CSM predictions.

\begin{figure}[htbp]
\begin{center} 
\includegraphics[width=7.5cm]{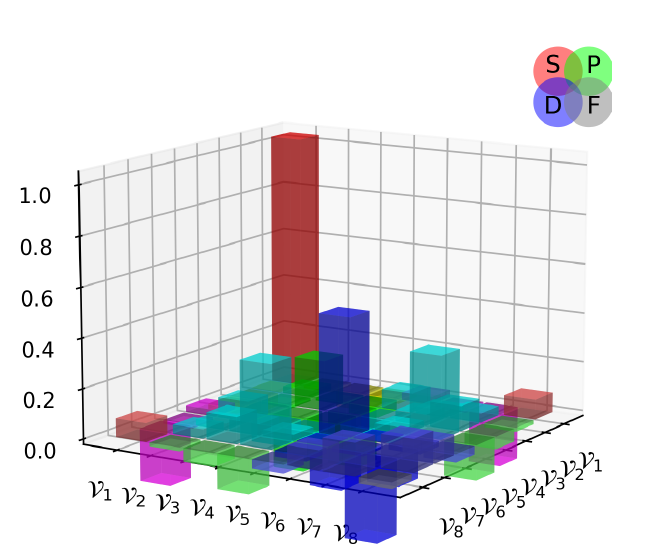}
\vspace{-1mm}
\caption{The contributions of quark$+$diquark configurations with orbital angular momenta $L$ and $L'$ in the canonical normalization 
constant of the $\Delta(1600)3/2^+$ quark core wave function computed within CSM in the resonance rest frame~\cite{Liu:2022ndb}. The
color code is shown in Fig.~\ref{d13 configurations}.} 
\label{delta_1600_configurations}
\end{center}
\end{figure}

The quark core structure of the $\Delta(1600)3/2^+$ in terms of the contributing quark$+$diquark configurations has been computed 
in Ref.~\cite{Liu:2022ndb}. It was found that nearly 100\% of the mass of this state is generated by the quark$+$diquark configuration 
with an axial-vector diquark of $J^P = 1^+$ and a quark in a relative $S$-wave in the resonance rest frame. The Chebyshev moment of 
this dominant configuration shows a clear zero crossing, providing evidence for a radial excitation. Hence, the underlying mass 
generation configuration for the $\Delta(1600)3/2^+$ quark core evaluated within the CSM is consistent with the SU(6) assignment of 
this state as the first radial excitation of three quarks in an  $S$-wave coupled to isospin 3/2. 

Studies of the contributions of quark$+$diquark configurations with orbital angular momenta $L$ and $L'$ in the canonical 
normalization constant for the resonance wave function in its rest frame again revealed a much more complex structure of this state 
as shown in Fig.~\ref{delta_1600_configurations}. The leading contribution arises from interference between $S$- and $F$-waves with 
sub-leading contributions from the $D$-wave and interference between $P$- and $D$-waves. As in the case of the $N(1520)3/2^-$, the 
mass of the $\Delta(1600)3/2^+$ does not have enough sensitivity to unravel the full complexity of the wave function
that can only be revealed in the studies of the $Q^2$-evolution of its electrocouplings. Confirmation of the CSM predictions on the 
$Q^2$-evolution of the $\Delta(1600)3/2^+$ electrocouplings by the experimental results obtained in this work validates the structure 
of its quark core evaluated within CSM under connection to the QCD Lagrangian.
 
\section{Conclusions and Outlook}
\label{concl_outlook}

A good description of the nine independent one-fold differential $\pi^+\pi^-p$ electroproduction cross sections off the proton has 
been achieved within the data-driven JM23 reaction model at $W$ from 1.41--1.66~GeV for $Q^2$ from 2.0--5.0~GeV$^2$. Comparable 
uncertainties have been achieved for the extracted resonant contributions in the model fits and the measured cross sections. The
resonance electrocouplings have been determined from the resonant contributions by employing a unitarized Breit-Wigner ansatz, allowing 
the restrictions imposed by a general unitarity condition on the resonant amplitudes in exclusive $\pi^+\pi^-p$ electroproduction to be
taken into account~\cite{Mokeev:2012vsa,Aitchison:1972ay}. The $\gamma_vpN^*$ electrocouplings of the $N(1440)1/2^+$, $N(1520)3/2^-$, 
and $\Delta(1600)3/2^+$ resonances have been determined from the $\pi^+\pi^-p$ electroproduction data for the first time for $Q^2$ 
from 2.0--5.0~GeV$^2$. 

The electrocouplings of the $N(1440)1/2^+$ and $N(1520)3/2^-$ determined in this work are in good agreement with the results determined
independently from the $\pi^+n$ and $\pi^0p$ electroproduction channels~\cite{Aznauryan:2009mx}. The consistency of the results from
independent studies of $\pi N$ and $\pi^+\pi^-p$ with completely different non-resonant contributions, supports the capabilities of 
the JM23 reaction model for the extraction of the resonance electrocouplings from $\pi^+\pi^-p$ electroproduction data. Furthermore,
consistent results on the electrocouplings of the $N(1440)1/2^+$, $N(1520)3/2^-$, and $\Delta(1600)3/2^+$ for $Q^2$ from 2.0--5.0~GeV$^2$
available from the independent fits of the $\pi^+\pi^-p$ electroproduction cross sections in three overlapping $W$ intervals with the 
contribution from the same resonance and different non-resonant amplitudes, solidifies the capability of the JM23 model to provide 
information on the resonance electrocouplings and their partial hadronic decay widths into the $\pi\Delta$ and $\rho p$ final states.

A successful description of the $\pi^+\pi^-p$ electroproduction cross sections achieved for $Q^2$ from 0.2--5.0~GeV$^2$ 
\cite{Mokeev:2008iw,Mokeev:2012vsa,Mokeev:2015lda} with the same $Q^2$-independent masses and total/partial hadronic decay widths 
for the $N(1440)1/2^+$, $N(1520)3/2^+$, and $\Delta(1600)3/2^+$, suggests that these nucleon excited states are produced in the 
$s$-channel of the $\gamma_v p$ interaction. 

The new results on the electrocouplings presented in this work extend the available information on the structure of $N^*$ states in the
mass range up to 1.6~GeV. Analyses of these results within the Argonne-Osaka coupled-channel approach~\cite{Suzuki:2009nj,Suzuki:2010yn,
Burkert:2017djo}, quark models~\cite{Obukhovsky:2011sc,Obukhovsky:2019xrs,Aznauryan:2018okk}, and within CSM under connection to the QCD
Lagrangian~\cite{Segovia:2015hra,Roberts:2021xnz}, have revealed the structure of these states as a complex interplay between an inner 
core of three dressed quarks and an external meson-baryon cloud. Studies of the electrocouplings over a broad range of $Q^2$ are critical
to reveal the structure of these states. At $Q^2 \lesssim 1$~GeV$^2$, the contribution from the meson-baryon cloud to the interaction 
with virtual photons at large distances is maximal. With increasing $Q^2$, photons of high virtuality penetrate the external meson-baryon
cloud and interact mostly with the core of three dressed quarks, elucidating the three-quark component in the structure of the $N^*$ 
states.

Continuum Schwinger methods~\cite{Segovia:2015hra,Carman:2023zke} have provided a successful description of the $N(1440)1/2^+$
electrocouplings for $Q^2$ from 2.0--5.0~GeV$^2$ by employing the momentum dependence of the dressed quark mass deduced from the
QCD Lagrangian. The CSM results are virtually the same as those obtained within the relativistic light-front quark model 
\cite{Aznauryan:2012ec,Aznauryan:2018okk}, which employed a phenomenological momentum-dependent dressed quark mass fit to the results 
on the nucleon electromagnetic elastic form factors. The predictions of the $Q^2$-evolution of the $\Delta(1600)3/2^+$ electrocouplings
made by CSM in 2019~\cite{Lu:2019bjs} with the same dressed quark mass function as described above have been confirmed by the results
obtained in this work from the $\pi^+\pi^-p$ electroproduction data. All of these studies demonstrate the relevance of dressed quarks 
with momentum-dependent running mass as the active degrees of freedom in $N^*$ structure.

The results of this work on the electrocouplings of the $N(1520)3/2^-$, as well as the previously available results from the studies of
$\pi N$ electroproduction~\cite{Aznauryan:2009mx}, provide a promising opportunity for CSM and other QCD-based approaches as they become
available to explore the relevance of dressed quarks to $N^*$s of different structure. In constituent quark models, this state is
treated as three quarks with orbital angular momentum $L = 1$ and belonging to the [70,1$^-$] SU(6) spin-flavor multiplet. CSM 
evaluations of the electrocouplings of this state are currently in progress.

In the near future we are planning to determine from the CLAS $\pi^+\pi^-p$ electroproduction data the electrocouplings of the most
prominent nucleon excited states in the mass range of $W$ from 1.6--2.0~GeV for $Q^2$ from 2.0--5.0~GeV$^2$. Analyses of these results 
provide additional promising opportunities for hadron structure theory to shed light on many facets of the dynamics in the strong QCD 
regime as more excited states of the nucleon with different structural features emerge from QCD. These studies also motivate the 
potential increase of the Jefferson Lab electron beam energy up to 22~GeV, which will offer a unique way to explore the full range 
of distances where $N^*$ structure emerges in the transition from the perturbative to the strongly coupled QCD regime
\cite{Accardi:2023chb}.

\begin{acknowledgments}

This work was supported in part by the U.S. Department of Energy (DOE) under Contract No. DE-AC05-06OR23177, the National Science
Foundation (NSF) under Grant PHY 1812382, the Physics Department of the University of South Carolina (USC) under NSF Grant PHY
10011349, Jefferson Science Associates (JSA), and the Skobeltsyn Nuclear Physics Institute and Physics Department at Lomonosov Moscow 
State University. ANHB is supported by
the DFG through the Research Unit FOR 2926 (project
number 409651613). We are grateful for theory support and constructive comments on the CSM approach from C. Chen, Z.-F. Cui, L. Liu, 
Y. Lu, C.D. Roberts, and J. Segovia, as well as on the quark models from I.G. Aznauryan, V.E. Lyubovitskij, I.T. Obukhovsky, and  
E. Santopinto.
\end{acknowledgments}

\nocite{*}
\newpage
\bibliography{References}
\end{document}